\pdfoutput=1 
\documentclass[journal,twoside,web]{ieeecolor}
\usepackage{generic}

\usepackage[english]{babel}
\usepackage{csquotes}

\usepackage{mathtools, amssymb}
\usepackage{subdepth} 
\usepackage{mathdots} 
\usepackage{graphicx}
\usepackage{float}
\usepackage{caption}
\usepackage{subcaption}

\usepackage[outline]{contour}

\usepackage{longtable}
\usepackage{booktabs}
\usepackage[l3]{csvsimple}
\usepackage{makecell}
\usepackage{pgfplotstable}

\usepackage[ruled,vlined]{algorithm2e}
\DontPrintSemicolon

\usepackage{tikz}
\usepackage{pgfplots}
\usetikzlibrary{math, positioning, shapes, arrows.meta, arrows, shapes.misc}
\usetikzlibrary{pgfplots.groupplots}
\pgfplotsset{compat=newest} 
\usepgfplotslibrary{statistics}
\usepgfplotslibrary{external}
\usepgfplotslibrary{polar}

\usepgfplotslibrary{colormaps}
\pgfplotsset{colormap/hot2}

\usepackage[backend=biber, style=ieee, isbn=false, doi=false, maxbibnames=3, minbibnames=1]{biblatex}

\bibliography{misc/radar_vs_references.bib}
\usepackage{jabbrv} 
\DeclareFieldInputHandler{booktitle}{
  \iffieldequalstr{entrytype}{inproceedings}
    {\def\NewValue{\JournalTitle{#1}}}
    {\def\NewValue{#1}}}
\makeatother

\makeatletter
\newcommand{\removelatexerror}{\let\@latex@error\@gobble}
\makeatother

\newcommand{\totaltau}{\tau_{pm}(l)} 
\newcommand{\tauheart}{\tau_{p}^{h}(l)} 
\newcommand{\tauresp}{\tau_{p}^{r}(l)} 
\newcommand{\tauvs}{\tau_{p}^{v}(l)} 
\newcommand{\tauloc}{\tau_{pm}^{d}} 

\newcommand{\distantenna}{\delta} 
\newcommand{\sigmat}{\bS_{l}}

\newcommand{\covmat}{\skew{1}\widehat{\bR}}
\newcommand{\covmattilde}{\skew{2}\tilde{\boldsymbol{R}}}

\newcommand{\textoverline}[1]{$\overline{\mbox{#1}}$}

\newcommand{\ba}{\boldsymbol{a}}

\newcommand{\bs}{\boldsymbol{s}}
\newcommand{\bw}{\boldsymbol{w}}

\newcommand{\blambda}{\boldsymbol{\lambda}}

\newcommand{\bA}{\boldsymbol{A}}
\newcommand{\bD}{\boldsymbol{D}}

\newcommand{\bH}{\boldsymbol{H}}
\newcommand{\bJ}{\boldsymbol{J}}

\newcommand{\bR}{\boldsymbol{R}}
\newcommand{\bS}{\boldsymbol{S}}
\newcommand{\bV}{\boldsymbol{V}}
\newcommand{\bX}{\boldsymbol{X}}

\newcommand{\vecop}{\text{vec}}

\pgfplotsset{
  discard if/.style 2 args={
    y filter/.code={
      \edef\tempa{\thisrow{#1}}
      \edef\tempb{#2}
      \ifx\tempa\tempb
      \def\pgfmathresult{inf}
      \fi
    }
  },
  discard if not/.style 2 args={
    y filter/.code={
      \edef\tempa{\thisrow{#1}}
      \edef\tempb{#2}
      \ifx\tempa\tempb
      \else
      \def\pgfmathresult{inf}
      \fi
    }
  }
}

\newlength{\figurewidth}
\newlength{\figureheight}

\definecolor{matlabblue}{rgb}{0 0.4470 0.741}
\definecolor{matlaborange}{rgb}{0.8500 0.3250 0.0980}
\definecolor{matlabyellow}{rgb}{0.9290 0.6940 0.1250}
\definecolor{matlabpurple}{rgb}{0.4940 0.1840 0.5560}
\definecolor{matlabgreen}{rgb}{0.4660 0.6740 0.1880}
\definecolor{matlablightblue}{rgb}{0.3010 0.7450 0.9330}
\definecolor{matlabred}{rgb}{0.6350 0.0780 0.1840}

\pgfplotscreateplotcyclelist{matlabcolor}{%
	matlabblue, mark=diamond*\\%
	matlaborange, mark=*\\%
	matlabyellow, mark=square*\\%
	matlabpurple, mark=triangle*\\%
	matlabgreen, mark=star\\%
	matlablightblue\\%
	matlabred\\%
}

\pgfplotsset{
/pgfplots/colormap={hot2}{[1cm]rgb255(0cm)=(0,0,0) rgb255(4cm)=(255,0,0) 
rgb255(7cm)=(255,255,0) rgb255(8cm)=(255,255,255)}}

\pgfplotstableread[col sep=comma,]{figures/tikz/Statistics_Person.csv}\PersonResultsTable
\pgfplotstableread[col sep=comma,]{figures/tikz/Statistics_Obstacle.csv}\ObstacleResultsTable

\pgfplotstableread[col sep=comma, columns={numPersons,MeanTPP,MeanFDP}]{figures/resultsTables/eval_round.csv}\resulttable

\usepackage{hyperref} 

\begin{document}
\title{Emergency Response Person Localization and Vital Sign Estimation Using a Semi-Autonomous Robot Mounted SFCW Radar}
\author{Christian A. Schroth, Christian Eckrich, \IEEEmembership{Student Member, IEEE}, Ibrahim Kakouche, Stefan Fabian, \\Oskar von Stryk, \IEEEmembership{Member, IEEE}, Abdelhak M. Zoubir, \IEEEmembership{Life Fellow, IEEE}, \\ and Michael Muma, \IEEEmembership{Senior Member, IEEE}
\thanks{Submitted for review to IEEE Transactions on Biomedical Engineering on 24. May 2023. This work was supported by LOEWE initiative (Hesse, Germany) within the emergenCITY centre. C. A. Schroth and A. M. Zoubir have been funded by DFG Project under grant ZO 215/19-1. M. Muma has been funded by the ERC Starting Grant ScReeningData (grant no. 101042407).}
\thanks{Christian A. Schroth, Christian Eckrich and Abdelhak M. Zoubir are with the Signal Processing Group, Technische Universität Darmstadt, Germany, (\{schroth, eckrich, zoubir\}@spg.tu-darmstadt.de). }
\thanks{Ibrahim Kakouche is with Laboratoire Systèmes Électroniques et Numériques, École Militaire Polytechnique, Bordj El-Bahri, Algeria, (brahim.kakouche@emp.mdn.dz).}
\thanks{Stefan Fabian and Oskar von Stryk are with the Systems Optimization and Robotics Group, Technische Universität Darmstadt, Germany (\{fabian, stryk\}@sim.tu-darmstadt.de).}
\thanks{Michael Muma is with the Robust Data Science Group, Technische Universität Darmstadt, Germany (michael.muma@tu-darmstadt.de).}}
\maketitle

\begin{abstract}
The large number and scale of natural and man-made disasters have led to an urgent demand for technologies that enhance the safety and efficiency of search and rescue teams. Semi-autonomous rescue robots are beneficial, especially when searching inaccessible terrains, or dangerous environments, such as collapsed infrastructures. For search and rescue missions in degraded visual conditions or non-line of sight scenarios, radar-based approaches may contribute to acquire valuable, and otherwise unavailable information. This article presents a complete signal processing chain for radar-based multi-person detection, 2D-MUSIC  localization and breathing frequency estimation. The proposed method shows promising results on a challenging emergency response dataset that we collected using a semi-autonomous robot equipped with a commercially available through-wall radar system. The dataset is composed of 62 scenarios of various difficulty levels with up to five persons captured in different postures, angles and ranges including wooden and stone obstacles that block the radar line of sight. Ground truth data for reference locations, respiration, electrocardiogram, and acceleration signals are included. The full emergency response benchmark data set as well as all codes to reproduce our results, are publicly available at \url{https://doi.org/10.21227/4bzd-jm32}. 
\end{abstract}

\vfill\break
\section{Introduction}
\label{sec:introduction}
\begin{figure}
	\centering
	\resizebox{0.9\columnwidth}{!}{\input{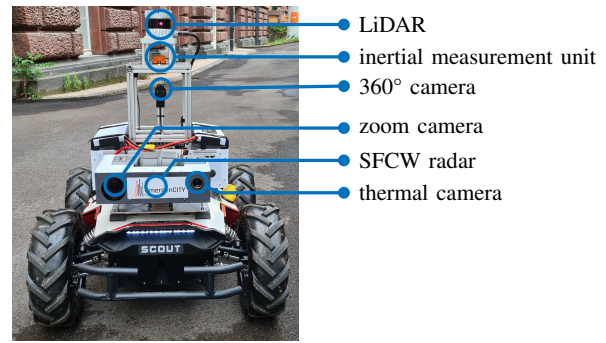}}
	\caption{emergenCITY\protect\footnotemark  robot 'Scout'}
	\label{fig:scout}
 \vspace{-5mm}
\end{figure}

\IEEEPARstart{L}{ocalizing} missing persons and monitoring their vital sings is an essential task in rescue missions, such as in a large scale natural disaster or a severe damage of critical infrastructure. The deployment of semi-autonomous exploration robots like the one shown in Fig.~\ref{fig:scout} is beneficial to enter environments that are dangerous or inaccessible to humans \cite{KruijffKorbayova.2021, Daun.2021, Hollick.2019}. Unlike humans, robots can, in principle, handle hazardous substances or extreme temperatures, do not become stressed or fatigued, and can be easily repaired or replaced. Commonly used technologies for autonomous mobile robot navigation strongly rely on visual sensing modalities like cameras or Light Detection and Ranging (LiDAR). Such sensors may be strongly impaired in a low-visibility environment, e.g., in the presence of severe smoke, dust, or fog, which regularly occur in crisis situations. Radar-based solutions for person localization \cite{Kim.2015, Starr.2014} or simultaneous localization and mapping (SLAM) \cite{Torchalla.2021} provide more reliable information. Furthermore, unlike thermal cameras, an appropriate radar system can sense the presence of humans through walls or other obstacles, which is of great interest for emergency responders \cite{Kozlov.2022, Kusmadi.2015, Leigsnering.2014, Amin.2011, Leigsnering.2011, Debes.2011, Debes.2009}.

\footnotetext{\url{https://www.emergencity.de/}}

The use of ultra wide band (UWB) radar for the localization of humans is an active area of current research. For example, Frequency Modulated Continuous Wave (FMCW) radar systems  \cite{Mercuri.2021b, Huang.2020, Feger.2009} and Stepped Frequency Continuous Wave (SFCW) radar systems \cite{Jia.2019, Tan.2018} have been used for location estimation in free space, and SFCW radar was also shown to be capable of through-wall localization \cite{Kozlov.2022, Kusmadi.2015, Leigsnering.2014, Leigsnering.2011}.

UWB radars are not only useful for estimating the location of humans, but are becoming increasingly popular to monitor vital signs (VS), like respiration and heart beat \cite{Wu.2023, Gupta.2022, Kakouche.2022, Kakouche.2021b, Mercuri.2021, Shyu.2019, Khan.2017, Ren.2015, Wang.2015}. In a clinical setting, radar-based VS estimation enables the monitoring of patients without physically attaching sensors to their skin \cite{Fioranelli.2023}, which is important for patients with burned skin, allergic reactions to the ECG electrode conduction gel \cite{Schellenberger.2020, Lim.2022} or for newborns \cite{Huang.2015}. Further application areas include the remote monitoring of senior citizens with a pre-existing illness \cite{Amin.2016}, in-home gait analysis \cite{Seifert.2019}, user verification \cite{Liu.2020} or work environment safety monitoring in a smart factory with the aim to detect an abnormal breathing of workers \cite{Islam.2022}. Recently, in \cite{Kakouche.2021c}, joint VS and location estimation of multiple persons for single-input multiple-output (SIMO) radars has been considered.

SFCW radar devices have been used to estimate the VS of a single person \cite{Nahar.2018b, Ren.2017, Quaiyum.2017, Ren.2015b} or multiple persons \cite{Paterniani.2023, Acar.2021, Su.2019}. In an emergency response, through-wall VS estimation is of particular interest, and only a few approaches have been proposed \cite{Li.2021b, Liu.2014}. Existing methods, however, usually only rely on a small number of measurements or even a single measurement as proof of concept and do not evaluate their proposed algorithm on a large and challenging dataset with a variety of different scenarios. Further, the selected scenarios often include simple or very idealized settings only. Additionally, the used datasets and codes are generally not made publicly available, which greatly limits the reproducibility of the proposed methods.

\emph{Main Contributions:} In this work, we propose, evaluate and make available a complete signal processing pipeline (flow chart, see Fig.~\ref{fig:flow_chart}) that includes clutter rejection, multi-person detection and 2D-MUSIC-based location estimation, vital-signs extraction and breathing rate estimation. Consistent with the envisioned application, the method is unsupervised and does not assume the availability of secondary information, such as environmental parameters or the number of subjects. We also provide an extensive benchmark dataset for multi-person detection, localization and vital sign estimation with a variety of different free-space, through-door and trough-wall scenarios. This benchmark dataset has been collected using a multiple-input multiple-output (MIMO) SFCW radar system that is integrated onto a semi-autonomous rescue robot, which can explore a disaster site in an autonomous or supervised manner to obtain a 3D map, including specific points of interest \cite{Daun.2021}. The diverse and challenging dataset with 62 different scenarios contains location information, as well as synchronously acquired vital signs and acceleration signals. Making available our codes and data allows the testing and development of new and existing algorithms in a wide range of different scenarios.  

The paper is organized as follows. Section~\ref{sec:signal_model} introduces the SFCW signal model and received range profile. Section~\ref{sec:method} details the proposed person detection, localization and vital sign estimation algorithms, followed by Section~\ref{sec:dataset}, which gives a detailed explanation of the provided benchmark dataset. Experimental results are presented in Section~\ref{sec:experiments}. Finally, a discussion is given in Section~\ref{sec:discussion} and conclusions are drawn in Section~\ref{sec:conclusion}.

\section{Signal Model}
\label{sec:signal_model}
\begin{figure}
	\centering
	\resizebox{0.73\columnwidth}{!}{\begin{tikzpicture}[scale=3]
	\tikzmath{
		\fo = 0.25;
		\df = 0.15; 
		\N = 7; 
		\f1 = \fo+\N*\df;
		\step = 0.25;
		\T =  \step*(\N+1);
	} 
	
	\coordinate (ymin) at (0, -0.1);
	\coordinate (ymax) at (0, 1.5);
	\coordinate (xmin) at (-0.1, 0);
	\coordinate (xmax) at (\step*\N+0.6,0);
	\coordinate (f0) at (0,\fo);
	\coordinate (f1) at (0,\fo+\df*\N);
	
	\draw [->] (ymin) -- (ymax);
	\draw [->] (xmin) -- (xmax);
	
	\draw[dashed] (-.05,\f1) -- (\T-\step, \f1);
	\draw[dashed] (-.05,\fo) -- (\T, \fo);
	\draw[] (0,\fo) foreach \x in {0,...,\N} {-|(\step*\x,\fo+\df*\x)};
	\draw[] (\step*\N,\f1) -- (\T,\f1);
	
	\node [left of = ymax, xshift = 0.5 cm] {$f$};
	\node [left of = f0, xshift = 0.5 cm] {$f_{0}$};
	\node [left of = f1, xshift = 0.5 cm] {$f_{K-1}$};
	
	\draw[dashed] (\step*3.5,\fo+\df*4) -- (\step*4, \fo+\df*4);
	\draw[dashed] (\step*3.5,\fo+\df*5) -- (\step*5, \fo+\df*5);
	\draw[<->] (\step*3.5,\fo+\df*4)  -- node[left]{$\Delta f$} (\step*3.5,\fo+\df*5)  ;
	
	\node [below of = xmax, yshift = 0.5cm] {$t$};
	
	\draw[dashed] (\step*4,\fo+\df*2) -- (\step*4, \fo+\df*3);
	\draw[dashed] (\step*5,\fo+\df*2) -- (\step*5, \fo+\df*4);
	\draw[<->] (\step*4, \fo+\df*2)  -- node[below]{$T$} (\step*5, \fo+\df*2)  ;
	
	\draw[<->] (\T,\fo) -- node[right]{$B$} (\T, \f1) ;
	
	\draw (\T,-.05) node[below]{$T_{K}$} -- (\T,.05);
	
\end{tikzpicture}}
	\caption{Single transmit waveform of an SFCW radar.}
	\label{fig:tx_signal}
    \vspace{-3mm}
\end{figure}
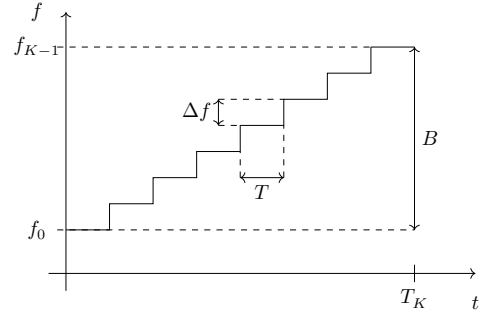
As shown in Fig.~\ref{fig:tx_signal}, an SFCW radar transmits a signal that is composed of $K$ sinusoidal signals of frequencies $f_{0},\dots,f_{K-1}$. Starting with $f_0$, the frequency is increased step-wise by $\Delta f$ to the maximal value of $f_{0} + (K-1) \Delta f$. The bandwidth is then given as $B = K \cdot \Delta f$. Each frequency tone is transmitted for a time interval $T$, such that the transmit time of a single waveform is given by $T_K = K\cdot T$. A single transmit waveform at time $t$ can thus be expressed as%
\begin{equation}
	x(t) = \sum_{k = 0}^{K-1} \textrm{rect} \left(\frac{t-\frac{T}{2}-kT}{T}\right) e^{-j 2 \pi (f_0 + k \Delta f) t}.
\end{equation}
Assuming that there are $P$ persons in a scene, the received signal at slow time index $l$ and receive antenna $m$ can be modeled by 
\begin{equation}
	s_{l}(t,m) =  \sum_{p = 1}^{P}\gamma_{pm} x\left(t - \totaltau \right) + \varepsilon_{m},
    \label{eqn:rxtime}
\end{equation}
where $\gamma_{pm}$ denotes the amplitude of the $p$th person, $ \varepsilon_{m}$ models clutter, and $m = 0,\dots, M_{r}-1$ where $M_{r}$ is the number of receive antennas. The time delay of the received signal at slow time index $l$, i.e., $\totaltau$  can be decomposed into two components, i.e.,
\begin{equation}
    \totaltau = \tauloc + \tauvs.
    \label{eqn:totaltau}
\end{equation}
The largest component in Eq.~\eqref{eqn:totaltau} is $\tauloc$, which relates to the distance between the radar and a person. It can be further decomposed, as it depends on the range $d_p$ and the angle $\theta_p$:
\begin{equation}
    \tauloc = \frac{1}{c} \left(2 d_p + m \distantenna \sin(\theta_p)\right),
\end{equation}
where $c$ is the speed of light and $\delta$ is the inter-antenna distance of the receiver uniform linear array (ULA), in which the first antenna ($m = 0$) is the reference. The second, much smaller component $\tauvs$ in Eq.~\eqref{eqn:totaltau} is the time delay caused by the respiration $\tauresp$ and the heart beat $\tauheart$ of Person $p$
\begin{equation}
    \tauvs = \tauresp +\tauheart,
\end{equation}
which changes over slow time but is commonly assumed to be constant over a single transmit waveform. After discretization and conversion of each frequency tone of the signal in Eq.~\eqref{eqn:rxtime} to the baseband, the normalized signal can be expressed as
\begin{equation}
	\tilde{s}_{l}(k,m) =  \sum_{p=1}^{P}e^{-j2 \pi (f_0+k \cdot \Delta f) \totaltau} + \tilde{\varepsilon}_{m}
	\label{eqn:rxfreq}
\end{equation}
for $k = 0,\dots,K-1$. Classically, for an SFCW radar, the range profile is calculated by computing an $N$-point inverse Fourier transform over the fast time frequency index $k$ \cite[p.~206~ff.]{Richards.2005}, \cite{Nguyen.2016}, i.e., for  $n = 0,\dots,N-1$
\begin{align}
    s_{l}(n,m) & = \frac{1}{K} \sum_{k = 0}^{K-1} \tilde{s}_{l}(k,m) e^{j2 \pi \frac{k n}{N}} \nonumber\\
            & = \frac{1}{K} \sum_{k = 0}^{K-1} \sum_{p=1}^{P}e^{-j2 \pi (f_0+k \cdot \Delta f) \totaltau}  e^{j2 \pi \frac{k n}{N}}\nonumber \\
            & = \frac{1}{K} \sum_{p=1}^{P} e^{-j2 \pi f_0 \totaltau} \sum_{k = 0}^{K-1}e^{j k \frac{2 \pi}{N} (n - N \Delta  f \totaltau) }.
    \label{eqn:inverseFourier}
\end{align}
Using the exponential sum formula \mbox{\cite[p.~138]{Kraus.1988}, \cite[p.~207]{Wehner.1995}}
 \begin{equation}
    \sum_{\rho = 0}^{\beta-1} e^{j\rho x} = \frac{\sin(\frac{\beta x}{2})}{\sin(\frac{x}{2})} e^{jx \frac{\beta-1}{2}}
\end{equation}
with $x = \frac{2\pi}{N} (n - N \Delta f \totaltau)$, Eq.~\eqref{eqn:inverseFourier} can be written as
\begin{align}
	s_{l}(n,m) = \frac{1}{K} & \sum_{p=1}^{P}  \frac{\sin\left(\pi \frac{K}{N}(n - N\Delta f \totaltau)\right)}{\sin\left(\frac{\pi}{N} (n - N\Delta f \totaltau)\right)} \times\nonumber \\
    & e^{-j2\pi f_0 \totaltau}  e^{j \pi\frac{K-1}{N} (n - N \Delta f \totaltau)}.
\end{align}
The range profile is obtained by taking the absolute value
\begin{align}
	r_{l}(n,m) = \left|s_{l}(n,m)\right|
	\label{eqn:rangeprofile}
\end{align}
with a maximum unambiguous range, a range resolution and range measurements in the interval \cite[p.~210]{Richards.2005} \cite[p.~50]{Nguyen.2016}
\begin{equation}
    d_\text{max} = \frac{c}{2 \Delta f}, \quad \Delta d = \frac{c}{2 B}, \quad \text{and} \quad \delta d = \frac{c}{2 \Delta f N},
    \label{eqn:dmax}
\end{equation} 
respectively. Here, $\Delta d$ denotes the minimum required distance between targets to distinguish two targets. By choosing $N > K$, the granularity of the range profile, represented by $\delta d$, is increased, as the envelope of the synthetic range profile is sampled more often \cite[p.~208]{Wehner.1995}.

\section{Proposed Localization and Vital Signs Estimation}
\label{sec:method}

This section presents a signal processing pipeline for multi-person detection, localization and vital sign estimation in emergency response using an SFCW radar that is mounted on a semi-autonomous rescue robot. The real-world applicability of the proposed method will be assessed in Sec.~\ref{sec:experiments} by evaluating the proposed method on a large and challenging real-world dataset (see Sec.~\ref{sec:dataset}) that is made publicly available along with all our codes to allow reproducibility and benchmarking. All steps of our proposed pipeline are detailed in the sequel. 

\subsection{Clutter Rejection}
\label{subsec:clutter}

The recorded radar signal $\tilde{s}_{l}(k,m)$ (see Eq.~\eqref{eqn:rxfreq}) contains clutter $\tilde{\varepsilon}$ caused by signal reflections from stationary objects located in the field of view, and a strong clutter component in the range bins close to the radar ($\leq 1$\,m) that is presumably caused by interference of the radar's internal components. 
In contrast to the signal of interest, the clutter components are stationary in slow time $l$. Therefore, we apply a sample moving average (SMA) filter\footnote{Different clutter rejection methods, e.g., using  mean subtraction and singular value decomposition (SVD) \cite{Liu.2017} were compared numerically. Based on these empirical results, which are not reported here due to page limitations, we settled for the SMA.}, i.e.,
\begin{align}
    s_{l}(k,m) = \tilde{s}_{l}(k,m) - \frac{1}{W_{\text{st}}}\sum^{l}_{i=l-W_{\text{st}}+1}\tilde{s}_{i}(k,m)
    \label{eqn:sma}
\end{align}
for $l = W_{\text{st}}, \dots ,L$, where $W_{\text{st}}$ denotes the slow time window length. The window must not be smaller than one breathing cycle to ensure the desired signal is maintained by the SMA filter.

\subsection{Virtual Antenna Array}
\label{subsec:virtual}

\begin{figure}
	\centering
	\resizebox{0.88\columnwidth}{!}{\begin{tikzpicture}[scale=0.75]
	\foreach \o in {0,4} {
		\draw[-] (\o-0.33,1) -- (\o+0.33,1);
		\draw[-] (\o-0.33,1) -- (\o,0.5);
		\draw[-] (\o+0.33,1) -- (\o,0.5);
		\ifnum \o = 0
		\draw[-] (\o,0) node[below]{$1$} -- (\o,1);
		\else
		\draw[-] (\o,0) node[below]{$2$} -- (\o,1);
		\fi
	}
	\draw[-] (0,0) -- node[midway, below]{Tx antennas}(4,0);
	\foreach \o [evaluate={\m=int(\o-6);}] in {6,...,9} {
		\ifnum \m < 1
		\draw[-] (\o,0) node[below]{$1$} -- (\o,1);
		\else
		\ifnum \m = 3
		\draw[-] (\o,0) node[below]{$M_{r}$} -- (\o,1);
		\else
		\draw[-] (\o,0) node[below, yshift=-0.5ex]{} -- (\o,1);
		\fi
		\fi
		\draw[-] (\o-0.33,1) -- (\o+0.33,1);
		\draw[-] (\o-0.33,1) -- (\o,0.5);
		\draw[-] (\o+0.33,1) -- (\o,0.5);
		\ifnum \o > 6
		\draw[-] (\o-1,0) -- (\o,0);
		\fi
	}
	\draw[-] (0,-1) -- (0,-1.3);
	\draw[-] (0,-1.15) -- (4,-1.15) node[midway,below] {$\delta_t$};
	\draw[-] (4,-1) -- (4,-1.3);
	\draw[-] (9,-1) -- (9,-1.3);
	\draw[-] (9,-1.15) -- (13,-1.15) node[midway,below] {$\delta_t$};
	\draw[-] (13,-1) -- (13,-1.3);
	\draw[-] (6,-1) -- (6,-1.3);
	\draw[-] (6,-1.15) -- (7,-1.15) node[midway,below] {$\distantenna$};
	\draw[-] (7,-1) -- (7,-1.3);
	\foreach \o [evaluate={\m=int(\o-6);}] in {10,...,13} {
		\ifnum \m > 6
		\draw[densely dotted] (\o,0) node[below]{$M$} -- (\o,1);
		\else
		\draw[densely dotted] (\o,0) node[below, yshift=-0.5ex]{} -- (\o,1);
		\fi
		\draw[densely dotted] (\o-0.33,1) -- (\o+0.33,1);
		\draw[densely dotted] (\o-0.33,1) -- (\o,0.5);
		\draw[densely dotted] (\o+0.33,1) -- (\o,0.5);
		\draw[densely dotted] (\o-1,0) -- (\o,0);
	}
	\foreach \o in {6,...,7} {
		\coordinate (A) at (\o,1+0.1);
		\coordinate (T) at (\o-0.5,2+0.1);
		\coordinate (T1) at (\o-0.5-0.5,2+1);
		\coordinate (S) at (\o+2/2.5,1/2.5+1+0.1);
		\draw[-] (A) -- (T);
		\draw[dotted] (T) -- (T1);
		\ifnum \o < 7
		\draw[-] (A) -- (S);
		\fi
	}
	
	\node[circle,draw,label=right:Person $p$] (c) at (5,3.5){};
	
	\draw[dashed] (7,1+0.1) -- (7,3);
	\draw (6+2/2.5,1/2.5+1+0.1) ++(-62:0.2) arc (-62:-152:0.2); 
	\draw (7,1+0.1) ++(90:1) arc (90:117:1) node[right, yshift=-1ex, xshift=2ex]{$\theta_p$}; 
\end{tikzpicture}}
	\caption{MIMO setup with two transmit antennas and $M_{r}$ physical receive antennas, which can be extended to $M$ virtual antennas.}
	\label{fig:virtualarray}
    \vspace{-3mm}
\end{figure}
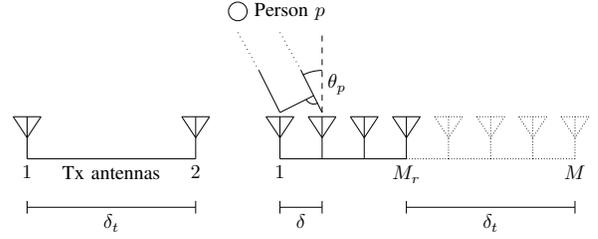

For a MIMO radar system, it is possible to increase the size of the receiving antenna array, using the phase differences at the different locations of the transmit and receive antennas. This virtual array has the advantage of an increased angular resolution, without deploying more physical antennas \cite{Rao.2018, Wang.2012, Ogawa.2018}. In particular, a linear receive array with $M_r$ antennas, which are equally spaced with distance $\distantenna$, can be extended to a virtual linear receive array of $M = M_r M_t$ antennas, where $M_t$ is the number of antennas in the linear transmit array with an inter-element spacing of $\delta_t = M_r \distantenna$ \cite{Li.2009}. In this work, we synthesize a virtual linear receive array as shown in Fig.~\ref{fig:virtualarray} with $M_r = 4$ and $M_t = 2$.

\subsection{2D-MUSIC Location Estimation and Spatial Smoothing}
\label{subsec:smoothing}

Commonly, the MUSIC algorithm is used in 1D Direction of Arrival (DoA) estimation \cite{Gentilho.2020}, but by adapting the steering vector to two dimensions, MUSIC becomes applicable for simultaneous DoA and range estimation \cite{Seo.2021, Zhang.2018, Garello.2008}, which is important to resolve situations where multiple persons occupy the same range bin (see Sec.~\ref{subsec:modelorder}). Following a similar strategy as \cite{Belfiori.2012, Ogawa.2018}, we propose a 2D-MUSIC location estimation algorithm that applies to the signal received by the extended virtual array shown in Fig.~\ref{fig:virtualarray}.  

The 2D-MUSIC algorithm is based on the singular value decomposition (SVD) of the covariance matrix of the received signal. With the resulting eigenvalues, the signal can be separated into a signal and noise subspace. In the case of correlated targets, the received signals are coherent, which leads to a rank deficient covariance matrix. Additionally, the dimension of the signal subspace will no longer be equal to the number of targets. Therefore, to decorrelate the signals, the forward and backward spatial smoothing (FBSS) method is applied \cite{Takao.1987, Pillai.1989, Odendaal.1994, Yi.2005, Shahimaeen.2019}. 


Starting with the received signal matrix $\sigmat$ of size $K \times M$ for slow time $l$, the signal from Eq.~\eqref{eqn:sma} is re-arranged, such that
\begin{align}
    \sigmat = &[\bs_{l}(0), \dots,  \bs_{l}(M-1)]  \\
        =& \begin{bmatrix}
        S_{0,0}^{l}  & \cdots & S_{0,M-1}^{l}\\
        \vdots  & \ddots & \vdots \\
        S_{K-1,0}^{l} & \cdots & S_{K-1,M-1}^{l}\\
    \end{bmatrix}
    \in \mathbb{C}^{K\times M}
    \label{eqn:S}
\end{align}
with $\bs_{l}(m) = \left[s_{l}(0,m), \dots,  s_{l}(K-1,m) \right]^{\top} \in \mathbb{C}^{K\times 1}$. 

In the remainder of this section, for visual clarity, we will drop the index $l$, but keep in mind that all following calculations are done for every slow time sample $l$. To process the data, we compute slices $\bS_{i,j}$ of size $W_K\times W_M$ of $\sigmat$ by extracting the values from the received signal matrix, i.e., 
\begin{align}
\bS_{i,j} = \begin{bmatrix}
        S_{i,j}  & \cdots & S_{i,j+W_M-1}\\
        S_{i+1,j}  & \cdots & S_{i+1,j+W_M-1}\\
        \vdots  & \ddots & \vdots \\
        S_{i+W_K-1,j} & \cdots & S_{i+W_K-1,j+W_M-1}\\
    \end{bmatrix}
    \label{eqn:Sij}
\end{align}
with $W_M \leq M $ and $W_K \leq K$. The sliding window moves from the \mbox{$(0,0)$th} element to the \mbox{$(K-W_K,M-W_M)$th} element and, in total, we obtain $W_{KM} = (K-W_K+1)\cdot(M-W_M+1)$ slices from $\sigmat$. All slices are re-arranged as follows
\begin{align}
    \bX  = & [\vecop(\bS_{0,0}), \vecop(\bS_{1,0}), \cdots, \vecop(\bS_{K-W_K,0}),\vecop(\bS_{0,1}),  \nonumber \\
    &  \cdots, \vecop(\bS_{K-W_K,M-W_M})] \in \mathbb{C}^{W_K W_M\times W_{KM}}
        \label{eqn:X}
\end{align}
from which the sample covariance matrix is computed as
\begin{align}
    \covmattilde = & \frac{1}{W_{KM}} \sum_{i = 0}^{K-W_K} \sum_{j = 0}^{M-W_M}\vecop(\bS_{i,j}) \vecop(\bS_{i,j})^{\mathsf{H}} \nonumber \\ 
     = &  \frac{1}{W_{KM}} \bX \bX^{\mathsf{H}} \in \mathbb{C}^{W_K W_M\times W_K W_M},
    \label{eqn:R}
\end{align}
where $\vecop (\cdot)$ denotes the column-wise stacking of the matrix and $(\cdot)^{\mathsf{H}}$ denotes the transpose conjugate. The final covariance matrix estimate $\covmat$ is obtained after forward backward averaging, that is,
\begin{equation}
    \covmat = \frac{1}{2} (\covmattilde  + \bJ \covmattilde^{\mathsf{H}} \bJ) \text{,} \quad
    \bJ = \begin{bmatrix}
            0 &  \cdots & 1\\
            \vdots  & \iddots  & \vdots \\
            1 & \cdots & 0\\
        \end{bmatrix}
    \label{eqn:Rfb}
\end{equation}
with the exchange matrix $\bJ \in \mathbb{R}^{W_K W_M\times W_K W_M}$.

From~\eqref{eqn:Rfb}, we compute the eigenvalue decomposition of $\covmat$, i.e.,
\begin{align}
    \covmat = & \bV \bD \bV^{\mathsf{H}},
    \label{eqn:svd}
\end{align}
where $\bD = \textrm{diag}(\lambda_1,\lambda_2,\dots,\lambda_{W_K W_M})$ with $\lambda_i$ being the $i$th largest eigenvalue and $\bV = [\bV_s, \bV_n] \in \mathbb{C}^{W_K W_M\times W_K W_M}$. The matrix $\bV$ is split into the noise subspace $\bV_n \in \mathbb{C}^{W_K W_M\times W_K W_M - P}$ and the signal subspace $\bV_s \in \mathbb{C}^{W_K W_M\times P}$, which is spanned by the first $P$ eigenvectors that correspond to the largest eigenvalues. In this application, $P$ is the number of persons in the radar field of view.

The angles $\hat{\theta}_{p} \in \{\hat{\theta}_{1},\dots,\hat{\theta}_{P}\}$ and distance estimates $\hat{d}_{p} \in\{\hat{d}_{1},\dots,\hat{d}_{P}\}$ are obtained from the $P$ largest maxima of the 2D-MUSIC pseudo spectrum that is given by 
\begin{equation}
    P_\textrm{MUSIC}(d,\theta) = \frac{1}{ \left|\ba^{\mathsf{H}}(d,\theta)\bV_n\bV_n^{\mathsf{H}} \ba(d,\theta) \right|}.
\end{equation}
Based on Eq.~\eqref{eqn:rxfreq}, the steering vector, which has to be stacked in a similar order as in Eq.~\eqref{eqn:X}, can be constructed as
\begin{align}
    \ba(d,\theta) & = \vecop \left(\bA_{d,\theta}(W_K, W_M)\right) \in \mathbb{C}^{W_K W_M\times 1},
\end{align}
where
\begin{align}
    &\bA_{d,\theta}(W_K, W_M) = \nonumber\\
    &\begin{bmatrix}
        a_{d,\theta}(0,0)  & \cdots & a_{d,\theta}(0,W_M-1)\\
        \vdots  & \ddots & \vdots \\
        a_{d,\theta}(W_K - 1,0) & \cdots & a_{d,\theta}(W_K-1,W_M-1)\\
    \end{bmatrix}
    \label{eqn:steeringMatrix}
\end{align}
is of size $W_K\times W_M$ with the individual entries given by
\begin{equation}
    a_{d,\theta}(k,m) = e^{-j2 \pi (f_0+k \Delta f)\frac{1}{c} (2 d + m \distantenna \sin(\theta) )}.
\end{equation}
Finding the maxima in $P_\textrm{MUSIC}(d,\theta)$, requires a grid search on $d \in [0, d_{\textrm{max}}]$ and $\theta \in [-\theta_{\textrm{max}}, \theta_{\textrm{max}}]$ with step sizes $\Delta \theta$ and $\Delta d$ resulting in a 2D-grid-size of  $\left\lfloor\frac{d_\textrm{max}}{\Delta d}+1\right\rfloor\times \left\lfloor\frac{2\theta_\textrm{max}}{\Delta \theta}+1\right\rfloor$.

\subsection{Model Order Estimation}
\label{subsec:modelorder}

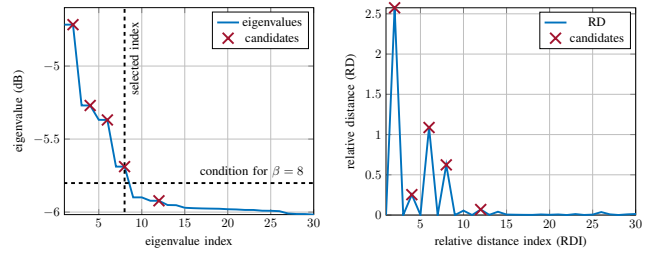
\begin{figure}
    \centering
        \subfloat[Eigenvalues sorted in descending order. The five candidate eigenvalues are marked with a red cross, the selected index and its corresponding condition are marked with dashed lines.] {\resizebox{0.48\columnwidth}{!}{\begin{tikzpicture}
        \begin{axis}[
            xlabel = eigenvalue index,
            ylabel = eigenvalue (dB),
            grid=both,
            xmin=1,
            xmax=30,
            legend style={anchor=north east},
            enlargelimits=false,
        ]
        \addplot+[thick,matlabblue,line width=1.5pt,no marks] table[x=x,y=lambda,col sep=comma] {figures/review/data_MOE_example_RD_lambda.csv};
        \addplot+[thick,matlabred,line width=1.5pt,only marks,mark=x,mark size=6pt] table[x=x, y=lambda_can, col sep=comma] {figures/review/data_MOE_example_candidates.csv};
        \addplot[mark=none, black, dashed, line width = 1.5, samples=2, domain = 0:30] {-5.8013} node[right,above,xshift=-1.7cm] {condition for $\beta = 8$};
        \addplot[mark=none, black, dashed, line width = 1.5, samples=2, smooth, domain=0:6] coordinates {(8,-6)(8,-4.6)} node[right, below, rotate=90, xshift=-1.4cm] {selected index};
        
        \legend{eigenvalues, candidates};
        \end{axis}
\end{tikzpicture}}%
        \label{fig:moe1}}
    \hfil
    \subfloat[Relative distances between two successive eigenvalues. The five dominant peaks are selected as candidates for the model order estimation algorithm.] {\resizebox{0.465\columnwidth}{!}{\begin{tikzpicture}
        \begin{axis}[
            xlabel = relative distance index (RDI),
            ylabel = relative distance (RD),
            grid=both,
            xmax=30,
            legend style={anchor=north east},
            enlargelimits=false,
        ]
        \addplot+[thick,matlabblue,line width=1.5pt,no marks] table[x=x,y=RD,col sep=comma] {figures/review/data_MOE_example_RD_lambda.csv};
        \addplot+[thick,matlabred,line width=1.5pt,only marks,mark=x,mark size=6pt] table[x=x, y=RD_can, col sep=comma] {figures/review/data_MOE_example_candidates.csv};
        \legend{RD, candidates};
        \end{axis}
\end{tikzpicture}}%
        \label{fig:moe2}}
    \caption{Visualization of the model order estimation procedure. Only the first 30 indices are shown.}
    \label{fig:moe}
\end{figure}

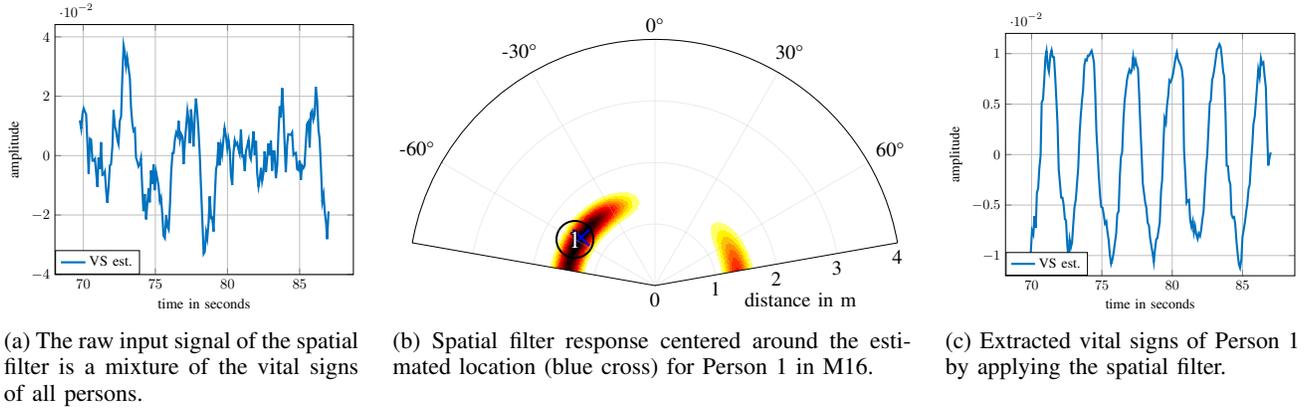
\begin{figure*}
    \centering
    \subfloat[The raw input signal of the spatial filter is a mixture of the vital signs of all persons.] {\resizebox{0.26\textwidth}{!}{\begin{tikzpicture}
    \begin{axis}[
            width = \figurewidth,
            height = \figureheight,
            xlabel = time in seconds,
            ylabel = amplitude,
            grid=both,
            cycle list name=matlabcolor,
            legend style={at={(0.001,0.001)},anchor=south west},
            axis background/.style={fill=white}
        ]
        \addplot+[thick,matlabblue,line width=1.5pt,no marks] table[x=x,y=etaAll,col sep=comma] {figures/review/eta-beamformer-ID16.csv};
        \legend{VS est.};
    \end{axis}
\end{tikzpicture}}%
        \label{fig:spatial1}}
    \hfil
    \subfloat[Spatial filter response centered around the estimated location (blue cross) for Person 1 in M16.] {\includegraphics[width=0.38\textwidth]{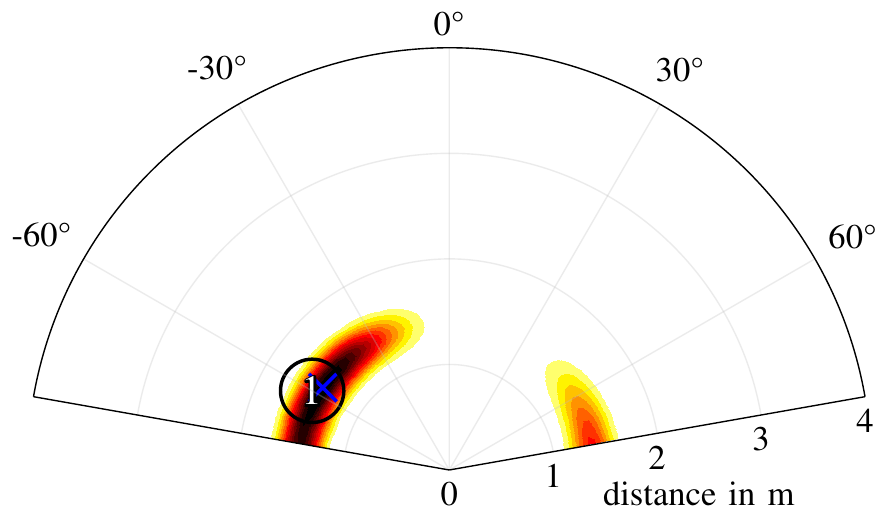}
    \label{fig:spatial2}}
    \hfil
    \subfloat[Extracted vital signs of Person 1 by applying the spatial filter.] {\resizebox{0.26\textwidth}{!}{\begin{tikzpicture}
    \begin{axis}[
            at={(1.2\linewidth,0)},
            width = \figurewidth,
            height = \figureheight,
            xlabel = time in seconds,
            ylabel = amplitude,
            grid=both,
            ymin=-0.012,
            ymax=0.012,
            cycle list name=matlabcolor,
            legend style={at={(0.001,0.001)},anchor=south west},
            axis background/.style={fill=white}
        ]
        \addplot+[thick,matlabblue,line width=1.5pt,no marks] table[x=x,y=est-p4,col sep=comma] {figures/review/eta-beamformer-ID16.csv};
        \legend{VS est.};
    \end{axis}
\end{tikzpicture}}%
        \label{fig:spatial3}}
    \caption{An illustration of the spatial filtering procedure for Person 1 in M16. The right plot, which shows the output of the spatial filter clearly demonstrates the extraction of the breathing signal.}
    \label{fig:spatial}
    \vspace{-3mm}
\end{figure*}

The MUSIC algorithm that was described in the previous section assumes the number of persons $P$ in the scene to be known. In real world scenarios, this is generally not the case. In the process of developing this method, different model order estimation methods were evaluated, including Akaike's Information Criterion (AIC), Minimum Description Length (MDL) \cite{Muma.2012, Wax.1985} and relative distance (RD) \cite{Jain.2012}. The investigated criteria are based on the eigenvalues of $\covmat$, that are computed based on Eq.~\eqref{eqn:svd} and collected into the vector $\blambda$. Some exemplary eigenvalues are shown in Fig.~\ref{fig:moe1}. According to our numerical evaluations, the best performance was achieved using the RD criterion, which is described in the sequel. 

The RD criterion is based on the relative distance between two successive eigenvalues, i.e., 
\begin{equation}
	\text{RD}(i)  =\frac{\lambda_{i} - \lambda_{i+1}}{\lambda_{i+1}}\quad \textrm{for} \quad i = 1,\dots,D.
\end{equation}
 From the relative distances $\text{RD}(i)$, the five largest candidate peaks with indices $\beta$ are extracted, as marked in Fig.~\ref{fig:moe2}. Starting with the largest index $\beta$, the first index $\beta$ satisfying the condition
\begin{equation}
	\lambda_{\beta} \ge \frac{\alpha}{D-\beta} \sum_{j=\beta+1}^{D} \lambda_{j}
	\label{eq:criterion}
\end{equation}
will be chosen and the model order estimate is given by 
\begin{equation}
    \hat{P} = \frac{\beta}{2}.
\end{equation}
The division by two comes from the fact that the data is complex-valued so that the eigenvalues occur in pairs, representing the real and imaginary part, and all $\text{RD}(i)$ with odd indices take values close to zero \cite{Jain.2012}. The parameter $\alpha \geq 1$ is a tuning variable. Due to spatial smoothing, we have observed an artificially introduced drop in the absolute value of the eigenvalues at index $D = \min \left(2 W_{KM}, W_M W_K\right)$, depending on the size of the smoothing window. As this drop disrupts the RD criterion, we will only consider eigenvalues with an index smaller than $D$. This does not impact the resulting model order estimate, as $D > P$.

\subsection{Spatial Filter for VS Extraction}
\label{subsec:spacialfilter}

A conventional non-adaptive spatial filter or beamformer \cite{Richards.2005} can be constructed for an estimated location $\{\hat{d}_p,\hat{\theta}_p\}$ by using the corresponding steering matrix
\begin{equation}
    \bA_{\hat{d}_p,\hat{\theta}_p}(K_0, M) \in \mathbb{C}^{K_0\times M}
\end{equation}
from Eq.~\eqref{eqn:steeringMatrix} with $K_0 \leq K$ (for details on $K_0$, see Eqs.~\eqref{eqn:deltafreq} and \eqref{eqn:K0} below). Side-lobe control can be introduced by multiplying the steering matrix with a distance and angular window function  $\bw_{K_0} \in \mathbb{R}^{K_0 \times 1}$ and $\bw_{M} ~\in ~\mathbb{R}^{M\times 1}$. The spatial filter is then given by
\begin{equation}
    \bH_{\hat{d}_p,\hat{\theta}_p} = (\bw_{K_0}\bw_M^{\top}) \circ \bA_{\hat{d}_p,\hat{\theta}_p}(K_0, M) ,
\end{equation}
where $\circ$ denotes the element-wise Hadamard product. Finally, the filter output is calculated by
\begin{equation}
    y_{\hat{d}_p,\hat{\theta}_p}(l) = \vecop\left(\bH_{\hat{d}_p,\hat{\theta}_p}\right)^{\mathsf{H}} \vecop\left(\sigmat^{K_0 \times M}\right)
    \label{eqn:applySpatialFilter}
\end{equation}
for every slow time sample $l$. The time delay $\tauvs$ caused by the vital signs of Person $p$ is encoded in the phase of $y_{\hat{d}_p,\hat{\theta}_p}(l)$ and can be related to the chest displacement according to
\begin{equation}
    \eta_{\hat{d}_p,\hat{\theta}_p}(l) = -\frac{c}{4 \pi f_c} \angle y_{\hat{d}_p,\hat{\theta}_p}(l)
\end{equation}
with the center frequency $f_{c} = f_{0} + \frac{K-1}{2}\Delta f$. The spatial filtering and vital signs estimation procedure is illustrated for the example of Person 1 in M16 in Fig.~\ref{fig:spatial}. Here, Fig.~\ref{fig:spatial1} shows the extracted phase signal, as given by the right-hand-side of Eq.~\eqref{eqn:applySpatialFilter}, while Fig.~\ref{fig:spatial3} shows the vital sign extracted at the estimated position as defined in the left-hand-side of Eq.~\eqref{eqn:applySpatialFilter}, that is marked as a blue cross in Fig.~\ref{fig:spatial2}. It should be noted that spatial aliasing can lead to ambiguity in the angular domain and the formation of grating lobes. This occurs if the spatial sampling frequency $1/\delta$ is too small, i.e., the distance between the antennas is too large. Hence, for an SFCW radar system, only the frequency steps $k$ can be used for the spatial filter that satisfy
\begin{equation}
    \delta \leq \frac{\lambda_k}{2} = \frac{c}{2f_0+2k\Delta f}.
    \label{eqn:deltafreq}
\end{equation}
Therefore, $K$ has to be restricted to 
\begin{equation}
    K_0\leq \left\lfloor\frac{c}{2\delta\Delta f}-\frac{f_0}{\Delta f}\right\rfloor \leq K,
    \label{eqn:K0}
\end{equation}
both for the construction of the spatial filter as well as for the signal matrix $S_l$. Further, to receive a continuous signal without phase jumps, the phase has to be unwrapped.

\subsection{Summary of the Proposed Algorithm}

\begin{figure}
	\centering
	\resizebox{0.95\columnwidth}{!}{\begin{tikzpicture}[scale=2.25]
	
	\node (RD) at (0,0) {Radar Data};
 
	\node [draw, below=0.5cm of RD, align=center]  (PreF) 
        {Clutter Removal \\ (Section~\ref{subsec:clutter})};

	\draw[-stealth] (RD.south) -- (PreF.north);

	\node [draw, below right=0.8cm and -0.2cm of PreF,	align=center]  (SMO1) 
        {Spacial Smoothing \\ for MUSIC \\ (Section~\ref{subsec:smoothing})};

 	\node [draw, below left=0.8cm and -0.2cm of PreF,	align=center]  (SMO2) 
        {Spacial Smoothing \\ for MOE \\ (Section~\ref{subsec:smoothing})};

	\draw[-stealth] (PreF.south) |- ($ (PreF)!0.4!(SMO1) $) -| (SMO1.north);
    \draw[-stealth] (PreF.south) |- node[below] {$\bS_l$} ($ (PreF)!0.4!(SMO1) $) -| (SMO2.north);

	\node [draw, below=1cm of SMO1, align=center]  (MUSIC) 
        {2D-MUSIC \\ (Section~\ref{subsec:smoothing})};

 	\node [draw, below=1cm of SMO2, align=center]  (MOE) 
        {Model Order \\ Estimation \\ (Section~\ref{subsec:modelorder})};

 	\draw[-stealth] (SMO1.south) -- (MUSIC.north) node[midway,left] {$\bR_\text{MUSIC}$};
    \draw[-stealth] (SMO2.south) -- (MOE.north) node[midway,left] {$\bR_\text{MOE}$};

	\node [draw, below=1cm of MUSIC, align=center]  (TargetS) 
        {Location Estimation};

	\draw[-stealth] (MUSIC.south) -- (TargetS.north) node[midway,left]{ $P_\text{MUSIC}$};

 	\draw[-stealth] (MOE.south) |- node[midway, left](P){$\hat{P}$} (TargetS.west);
 
	\node [draw, below=1cm of TargetS, align=center]  (SF) 
        {Spacial Filter and\\ Vital Sign Extraction \\ (Section \ref{subsec:spacialfilter})};

    \draw[-stealth] (TargetS.south) -- (SF.north) node[midway,left](Loc){$\{\hat{d}_p, \hat{\theta}_p\}$};
	\draw[-stealth] (PreF.south) |- ($ (PreF)!0.4!(SMO1) $) -| (2.2,-4) |- (SF.east);
    
	\node [draw, below=1cm of SF, align=center]  (Track) {Person Tracking};

	\node [draw, below=0.5cm of Track, align=center]  (VS) {Vital Sign Estimation \\ per Person};
	
	
	\draw[-stealth] (SF.south) -- (Track.north)
		node[midway,left]{$\eta_{\hat{d}_p,\hat{\theta}_p}(l)$};
	
	\draw[-stealth] (Track.south) -- (VS.north);

	\node [below=0.5cm of VS, align=center]  (VSEst) {Vital Signs};
    \draw[-stealth] (VS.south) -- (VSEst.north);

\end{tikzpicture}}
	\caption{Flow chart of the processing steps to extract vital signs. These steps have to be repeated for every slow time segment.}
	\label{fig:flow_chart}
\end{figure}
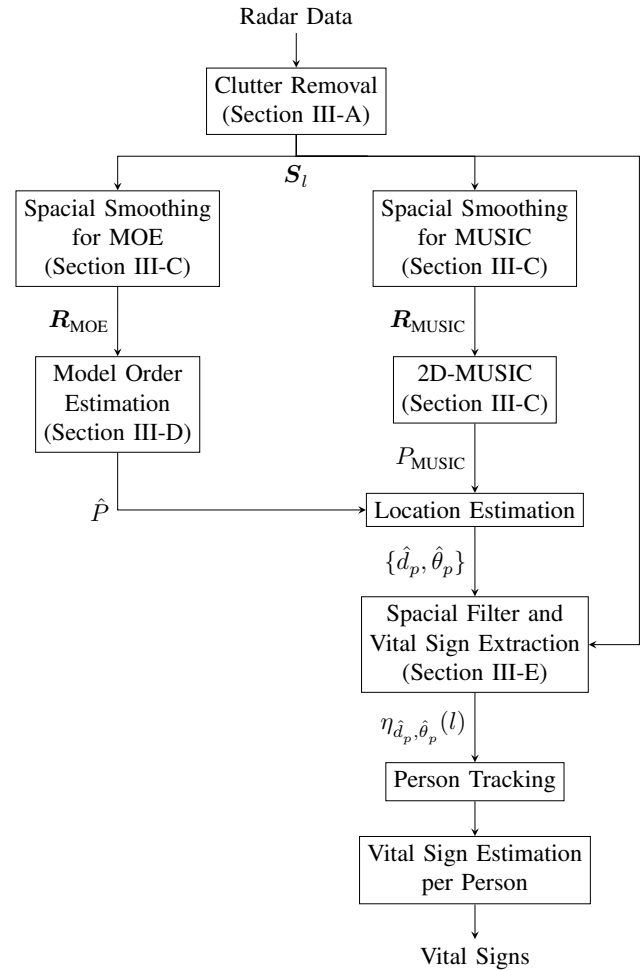

The data is sliced into segments of $L_{\text{st}} \leq L$ slow time samples. Then, the algorithm is applied to every segment as follows:
\begin{enumerate}
    \item \textit{Clutter removal:} Remove the clutter as detailed in Section~\ref{subsec:clutter}.
    \item \textit{Spatial smoothing for the MUSIC algorithm:} Apply spatial smoothing with $W_{KM} = W_{\text{MUSIC}}$ and calculate the MUSIC pseudo spectrum $P_\textrm{MUSIC}(d,\theta)$ as detailed in Section~\ref{subsec:smoothing}. Sum up the MUSIC pseudo spectra of all previous segments and the current spectrum to suppress the occurrence of spurious false detections.
    \item \textit{Spatial smoothing for model order estimation:} According to our experiments, to estimate the model order, it is beneficial to choose a different size for the spatial smoothing window, i.e., $W_{KM} = W_{\text{MOE}}$.
    \item \textit{Model order estimation:} Estimate the number of persons $P$ in the scene using the RD-algorithm as detailed in Section~\ref{subsec:modelorder} with the estimated smoothed covariance matrix from the previous step.
    \item \textit{Estimate locations:} Estimate the locations $\{d_p,\theta_p\}$ for $p=1,\dots,\hat{P}$ by extracting the highest peaks from the MUSIC spectrum. If there are multiple closely spaced peaks ($< 30$\,cm), then these peaks are grouped together and an additional peak is selected until $\hat{P}$ locations are found.
    \item \textit{Vital signs extraction:} Extract the vital signs $\eta_{\hat{d}_p,\hat{\theta}_p}(l)$ for every slow time sample from the estimated locations from the previous step with a spacial filter as detailed in Section~\ref{subsec:spacialfilter}.
    \item \textit{Person tracking:} As not all persons are found in every segment, 
    the detected persons have to be tracked from segment to segment. In particular, the same label to a detected person is assigned if their location is less than 25\,cm away from the location of a previously detected person\footnote{Note that a more sophisticated tracker is to be used when considering a non-stationary dataset.}. 
    \item \textit{Vital sign estimation:} For every detected person, estimate the breathing frequency by searching for the highest peak in the estimated spectrum.
\end{enumerate}
A flow chart of the algorithm is given in Fig.~\ref{fig:flow_chart}. The code is available at \url{https://github.com/schrchr/radar-vitals-estimation}.

\section{Emergency Response Benchmark Dataset}
\label{sec:dataset}

\subsection{Walabot Radar}
The benchmark dataset\footnote{\url{https://doi.org/10.21227/4bzd-jm32}} was collected with an SFCW MIMO radar system ('Walabot') by the manufacturer vayyar. This type of radar has been already used for soil moisture estimation \cite{Uthayakumar.2022, Niu.2020}, through-wall pose estimation \cite{Meng.2019} and for respiration monitoring \cite{Han.2022}. Detailed technical specifications are summarized in Table~\ref{tab:specs}. 

\begin{figure}
	\centering
	\resizebox{\columnwidth}{!}{\input{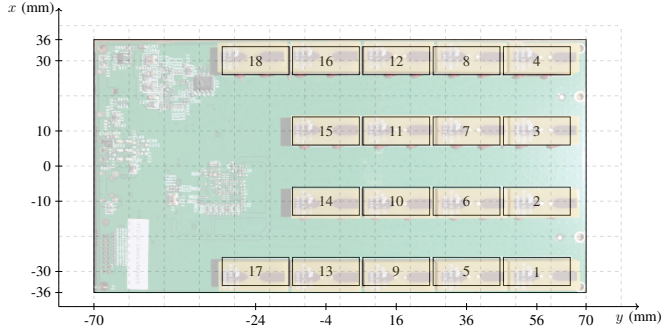}}
	\caption{Physical positions of the antennas, where the background shows the PCB of the radar. Note that $x$ and $y$ are denoted as given in the official documentation of 'Walabot'.}
	\label{fig:antenna_positions}
    \vspace{-3mm}
\end{figure}

The radar array is comprised of 18 dipole antennas in a grid layout with an inter-element spacing of $\delta =2$\,cm as shown in Fig.~\ref{fig:antenna_positions}. We synthesize a virtual linear receive array with $\delta_{t} = 8$\,cm as defined in Section~\ref{subsec:virtual} by performing a spatial convolution of the transmit and receive antennas indexed by $\{1, 17\}$ and $\{2,6,10,14\}$ in the 'Walabot' radar, respectively. As the 'Walabot' radar uses separation in the time domain and outputs the signal for each transmit and receive antenna (antenna pair) separately, the received signal for the virtual antenna array can be constructed by appending the signals associated with the second transmit antenna to those associated with the first one. 

The 'Walabot' offers three different scan profiles. The dataset was collected using the \verb|SENSOR| profile, which allows for a higher resolution, but has a slower capture rate\footnote{\url{https://api.walabot.com/_features.html}} compared to the other available settings. 

\begin{table}
	\centering
	\begin{tabular}{r l l} 
		\toprule
		$K$ & 137 & Number of frequency steps \\ 
        $K_0$ & 96 & Number of frequency steps to avoid ambiguity\\ 
		$N$ & 8192 & Number of sampling points in the range profile \\ 
		$T_K$ & 14.3\,$\mu$s & Sweep time \\ 
		$f_0$ & 6.3\,GHz & Start frequency \\ 
		$f_{K-1}$ & 8\,GHz & Stop frequency \\ 
		$B$ & 1.7\,GHz & Bandwidth \\ 
		$\delta$ & 2\,cm & Inter-antenna distance \\
        $d_{\text{max}}$ & 12.08\,m & maximum unambiguous range\\
        $\Delta d$ & 8.8\,cm & range resolution of two targets\\
		\bottomrule
	\end{tabular}
	\caption{Technical specifications of the 'Walabot' radar.}
	\label{tab:specs}
 \vspace{-3mm}
\end{table}

Only specific antennas (i.e., 1, 4, 17, and 18) can be used as transmit antennas. This leads to overall 40 possible antennas pairs for the \verb|SENSOR| profile, which are listed in Table~\ref{tab:AntennaPairs}. An overview of the necessary steps to conduct a data acquisition are given in Algorithm~\ref{alg:walabot_pseudocode}. More specifically, the function \textit{Trigger()} starts a measurement for all 40 antenna pairs, which can be accessed individually by calling the function \textit{GetSignal(tx, rx)}. The function \textit{GetSignal(tx, rx)} returns a range profile of length $N$ for each antenna pair at a slow time sample $l$. The slow time sampling frequency $f_{\text{st}}$ depends on the computational power of the measurement computer. In our setup, the slow time sampling frequency varied from 10\,Hz to 11\,Hz, which is sufficient for respiration (typically 0.25\,Hz to 0.4\,Hz) and heart beat (typically 0.9\,Hz to 1.6\,Hz) estimates \cite{Paterniani.2023}. For each measurement, the slow time sampling frequency can be computed using the provided slow time vector, which contains the exact time point of each trigger event.

\begin{table}
	\centering
	\begin{tabular}{ r l }
        \toprule
		Transmit  & Receive antenna \\
		\hline
		1& 2, 3, 6, 7, 10, 11, 14, 15, 4, 8, 12, 16, 18 \\
		4& 2, 3, 6, 7, 10, 11, 14, 15\\
		17& 2, 3, 6, 7, 10, 11, 14, 15, 8, 12, 16\\
		18& 2, 3, 6, 7, 10, 11, 14, 15\\
        \bottomrule
	\end{tabular}
	\caption{Antenna pairs of \texttt{SENSOR} profile.}
	\label{tab:AntennaPairs}
\end{table}

\begin{figure}
\centering
\begin{minipage}{0.9\columnwidth}
\removelatexerror
\begin{algorithm}[H]
	    ConnectAny()\\
	    SetProfile(\verb|SENSOR|)\\
	    SetDynamicImageFilter(\verb|FILTER_TYPE_NONE|)\\
        Start()\\
		\For{$l = 1,\dots,L$}
	    {%
           Trigger()\\
            \For{all antenna pairs}
            {%
                GetSignal(tx antenna, rx antenna)\\
            }
		}
        Stop()\\
        Disconnect()\\
        Clean()\\
	\caption{Pseudocode of data acquisition.}
	\label{alg:walabot_pseudocode}
\end{algorithm}
\end{minipage}
\end{figure}

\subsection{Measurement Setup and Protocol}
\subsubsection{Radar Data}
\begin{figure}
	\centering
	\resizebox{0.75\columnwidth}{!}{\begin{tikzpicture}[scale=1.2]

	\begin{scope}
		\clip (-3.5,0) rectangle (3.5,3.5);
		\draw (0,0) circle(1);
		\draw (0,0) circle(1.5);
		\draw (0,0) circle(2);
		\draw (0,0) circle(2.5);
		\draw (0,0) circle(3);
		\draw (0,0) circle(3.5);
	\end{scope}
	\draw (0,0) -- ( {3.5*sin(60)} , {3.5*cos(60)} ) node [pos=1.05, above] {60°};
	\draw (0,0) -- ( {3.5*sin(30)} , {3.5*cos(30)} ) node [pos=1.03, above] {30°};
	\draw (0,0) -- ( {3.5*sin(45)} , {3.5*cos(45)} ) node [pos=1.02, above] {45°};
	\draw (0,0) -- ( {3.5*sin(-45)} , {3.5*cos(-45)} ) node [pos=1.02, above] {-45°};
	\draw (0,0) -- ( {3.5*sin(-30)} , {3.5*cos(-30)} ) node [pos=1.03, above] {-30°};
	\draw (0,0) -- ( {3.5*sin(-60)} , {3.5*cos(-60)} ) node [pos=1.05, above] {-60°};
	\draw (0,0) -- ( {3.5*sin(0)} , {3.5*cos(0)} ) node [pos=1, above] {0°};
	\draw (-3.5,0) -- (3.5,0) node [pos=0.65, below] {1\,m} node [pos=0.79, below] {2\,m} node [pos=0.93, below] {3\,m};
	
	\foreach \o  in {-0.3,-0.1,0.1,0.3} {
		\draw[-, thick] (\o-0.06,-0.2) -- (\o+0.06,-0.2);
		\draw[-, thick] (\o-0.06,-0.2) -- (\o,-0.35);
		\draw[-, thick] (\o+0.06,-0.2) -- (\o,-0.35);
		\draw[-, thick] (\o,-0.2) -- (\o,-0.45);
	}
    \draw[-, thick] (-0.3,-0.45) -- node[below] {rx} (0.3,-0.45);

    \node[circle, draw, inner sep=0pt, minimum size = 20.4, fill, text=white] (d) at ({1.5*sin(-60)}, {1.5*cos(-60)}) {\textbf{1}};
    \node[circle, draw, inner sep=0pt, minimum size = 20.4, fill, text=white] (d) at ({2*sin(0)}, {2*cos(0)}) {\textbf{2}};
    \node[circle, draw, inner sep=0pt, minimum size = 20.4, fill, text=white] (d) at ({2*sin(-30)}, {2*cos(-30)}) {\textbf{3}};
    \node[circle, draw, inner sep=0pt, minimum size = 20.4, fill, text=white] (d) at ({2*sin(45)}, {2*cos(45)}) {\textbf{4}};
    \node[circle, draw, inner sep=0pt, minimum size = 20.4, fill, text=white] (d) at ({3*sin(30)}, {3*cos(30)}) {\textbf{5}};

\end{tikzpicture}}
	\caption{Grid view for M16.}
	\label{fig:field_of_view}
    \vspace{-3mm}
\end{figure}
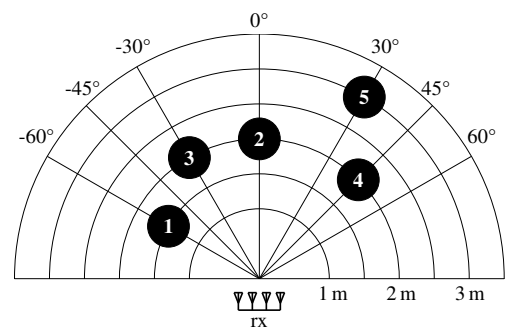

The 'Walabot' radar was integrated into the front part of the emergenCITY 'Scout' robot (behind the emergenCITY logo, see Fig.~\ref{fig:scout}) at a height of 48\,cm and with an elevation of 0°. The floor was marked with a grid as shown in Fig.~\ref{fig:field_of_view}, with azimuth angles $\theta$ = \{-60°, -45°, -30°, 0°, 30°, 45°, 60°\} and radial distances $d$ = \{1\,m, 1.5\,m, 2\,m, 2.5\,m, 3\,m, 3.5\,m\}, enabling the defined positioning of the participants. 

\begin{figure}
	\centering
	\resizebox{0.9\columnwidth}{!}{\input{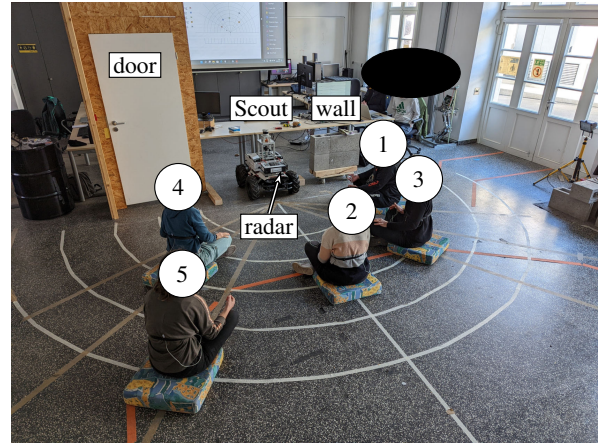}}
	\caption{Recording scene of M16.}
	\label{fig:measurement_setup}
     \vspace{-3mm}
\end{figure}

Measurements with different obstacles were implemented as follows; (i) a wooden door with a thickness of 3.6\,cm was placed 21\,cm in front of the radar; (ii) a stone wall of hollow cinder blocks with a height of 69\,cm and a thickness of 17.2\,cm was placed 7\,cm in front of the radar. The obstacles completely blocked the line of sight between the radar and the participants. An exemplary measurement setup is shown in Figs.~\ref{fig:measurement_setup} and \ref{fig:field_of_view}. Fig.~\ref{fig:measurement_setup} shows a photo with the markings for the individual participants while Fig.~\ref{fig:field_of_view} represents the location information in a grid view.

\subsubsection{Reference Data}
Synchronous reference data for each person was collected using a piezoelectric respiration sensor, an electrocardiograph (ECG) sensor and an accelerometer. All three sensors were connected to a BITalino\footnote{\url{https://www.pluxbiosignals.com/pages/bitalino}} device, which is an open-source biosignal platform by the manufacturer 'PLUX'. The data was recorded with the software OpenSignals~(r)evolution.

The respiration sensor measures the expansion and contraction of the participants chest with a piezoelectric elastic band. The measured changes of circumference of the chest linearly relate to the displacement of the chest, as measured by the radar, but the output voltage of the sensor cannot be traced back to a specific circumference. Nevertheless, the oscillations of the measured signal can be used as a reference signal for the respiration rate of the participants.

The ECG sensor is used as a reference signal to assess the heart rate of the participants. Two of the electrodes are placed over the thumb's and ring finger's metacarpals on the right hand, while the third electrode is placed over the thumb's metacarpals on the left hand.

As movements of the participants may cause artifacts in the measured radar and reference signals, an accelerometer sensor was attached on the back of each of the participants to capture these potential distortions.

\subsection{Dataset}
\subsubsection{Composition}
The benchmark dataset is composed of 121 files with 62 radar measurements and 59 reference vital parameter measurements. The radar measurements with IDs 0, 0-1 and 0-2 have no corresponding reference file as they are empty background measurements. Each measurement containing a person has a length of 2000 slow time samples (i.e., about 160\,s), while the empty room settings have a length of 200 slow time samples (i.e., about 16\,s). An overview of the measurements is given in Table~\ref{tab:measurements_summary} and a detailed list of all measurements containing the type of obstacle, and the positions and postures of the subjects is provided in the supplementary material.

Positions of the participants were chosen to ensure different levels of difficulties. Recording IDs M1 to M11 contain measurements of a single participant with different postures, angles, ranges and obstacles. The next set of measurements with IDs M12 to M26, are the most challenging free space measurements as they contain measurements of four or five participants in different ranges and angles, including multiple participants in the same range. Starting with measurement ID M27 all measurements are performed three times, for free space, with wall and with door. This set contains measurements with two and three participants in various arrangements. Scenarios in which one person is blocking the line of sight of another person were not recorded.  

Gender and age differed among the participants, but this information was not recorded and is not part of the dataset. The ethics commission of 'Technische Universität Darmstadt' (EK 30/2023, date of approval: 10. May 2023) gave their consent in recording and publication of the dataset.

\begin{table}
	\begin{center}
		\begin{tabular}{ c c c }
            \toprule
			\#recordings & \#persons & obstacle \\ 
			\hline
			7 & 0-1 & free space \\  
			3 & 0-1 & door \\    
			4 & 0-1 & wall \\    
			15 & 4-5 & free space \\    
			\hline
			11 & 2-3 & free space \\  
			11 & 2-3 & door \\  
			11 & 2-3 & wall \\    
            \bottomrule
		\end{tabular}
	\end{center}
	\caption{Overview of conducted experiments.}
	\label{tab:measurements_summary}
\end{table}

\subsubsection{Generating the Dataset from the Measurements}


The function \textit{GetSignal(tx, rx)} returns a range profile $r_{l}(n)$ of length $N = 8192$ for a single antenna pair as defined in Eq.~\eqref{eqn:rangeprofile}. This leads to a huge amount of data and, as explained in Section~\ref{sec:signal_model}, it does not contain any additional information. Hence, we downsample the signal to its original length of 137 samples. Starting by downconverting the signal to the baseband
\begin{equation}
    r_{l,\text{base}}(n) = r_{l}(n) \cdot e^{-j2 \pi  f_{c} n / f_{s}^{\text{ft}}}
\end{equation}
with $n = 0,\dots,8191$, the fast time sampling frequency $f_{s}^{\text{ft}} = 102.4$\,GHz and the center frequency $f_{c} = 7.15$\,GHz. Subsequently, we perform an FFT over fast time~$n$ and downsample the signal in the frequency domain by symmetrically removing all samples outside the frequency range $[-B/2, B/2] = [-0.85, 0.85]$\,GHz. Repeating these steps for all antenna pairs and slow time samples, leads to $\tilde{s}_{l}(k,m)$ as defined in Eq.~\eqref{eqn:rxfreq}. 


\section{Experimental Results}
\label{sec:experiments}

This section presents the experimental results. We start by introducing the used performance metrics and tuning parameters, followed by the detection, localization and vital sign estimation results that are obtained by applying the proposed algorithm to the benchmark dataset. 

\subsection{Performance Metrics}
To evaluate the performance of the proposed algorithm, we start by matching the reference location $\{d_p,\theta_p\}$ with the estimated location $\{\hat{d}_p,\hat{\theta}_p\}$. Two locations are considered to be matched if their Euclidean distance is smaller than $d_{\text{match}} = 0.3$\,m.
The first performance metric we use is the circular position error. It is calculated as the Euclidean distance between all matched locations. Based on the matches, we calculate the number of missed detections $P_{\text{MD}}$ (persons present, but not detected) and the number of false detections $P_{\text{FD}}$ (persons detected, but not present). These estimated values are related to the true number of persons according to $P = \hat{P} + P_{\text{MD}} - P_{\text{FD}}$. To allow for a comparison of different scenarios, the true positive proportion (TPP) 
\begin{equation}
    \text{TPP} = \frac{P - P_{\text{MD}}}{P},
\end{equation}
where values closer to one indicate a better performance and the false detection proportion (FDP)
\begin{equation}
    \text{FDP} = \frac{P_{\text{FD}}}{\max\{1,\hat{P}\}},
\end{equation}
where values closer to zero indicate a better performance, are calculated. FDP and TPP averaged over the different scenarios are denoted as \textoverline{FDP} and \textoverline{TPP}, respectively.

\subsection{Choice of Tuning Parameters}

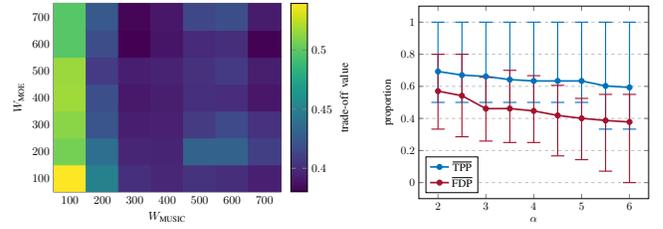
\begin{figure}
    \centering
    \subcaptionbox{Result of the trade-off $0.3 \cdot \overline{\text{FDP}} + 0.7 \cdot (1 - \overline{\text{TPP}})$ for a fixed $\alpha = 3$. Lower is better. \label{fig:w}}[.5\columnwidth]{\resizebox{.52\columnwidth}{!}{\begin{tikzpicture}
    \begin{axis}[
        enlargelimits=false,
        colorbar,
        ymin=50,
        ymax=750,
        xmin=50,
        xmax=750,
        ytick={100,200,300,400,500,600,700},
        xlabel=$W_{\text{MUSIC}}$,
        ylabel=$W_{\text{MOE}}$,
        colormap name={viridis},
        colorbar style={ylabel=trade-off value}
    ]
    \addplot [matrix plot*, mesh/cols=7, point meta=explicit] table [col sep=comma, meta=C] {figures/review/eval_W_alpha-3.csv};
    \end{axis}
\end{tikzpicture}}}
    \hfill
    \subcaptionbox{\textoverline{FDP} and \textoverline{TPP} over different values of $\alpha$ for fixed $W_{\text{MOE}} = 600$ and $W_{\text{MUSIC}} = 700$. The error bars indicate the 0.25 and 0.75 quantiles.\label{fig:alpha}}[.48\columnwidth]{\resizebox{0.41\columnwidth}{!}{\begin{tikzpicture}
        \begin{axis}[
        	width = \figurewidth,
	        height = \figureheight,
            xlabel = $\alpha$,
            ylabel = proportion,
            ymajorgrids,
            grid style=dashed,
            ymin=0,
            ymax=1,
            enlarge x limits=0.1,
            enlarge y limits=0.1,
            cycle list name=matlabcolor,
            legend pos = south west,
        ]
        \addplot+[thick,matlabblue,line width=1.5pt,mark=*,error bars/.cd, error bar style={line width=1pt}, error mark options={rotate=90,matlabblue,mark size=6pt,line width=1pt}, y dir=both, y explicit] table[x=alpha,y=TPP,col sep=comma, y error plus=QuanHiTPP, y error minus=QuanLoTPP] {figures/review/review_eval_alpha_Wmoe-600_Wmusic-700_numWin-10_Ncov-10.csv};
        
        \addplot+[thick,matlabred,line width=1.5pt,mark=*,error bars/.cd, error bar style={line width=1pt}, error mark options={rotate=90,matlabred,mark size=6pt,line width=1pt}, y dir=both, y explicit] table[x=alpha,y=FDP,col sep=comma, y error plus=QuanHiFDP, y error minus=QuanLoFDP] {figures/review/review_eval_alpha_Wmoe-600_Wmusic-700_numWin-10_Ncov-10.csv};
        \legend{\textoverline{TPP}, \textoverline{FDP}};
        \end{axis}
\end{tikzpicture}}}
    \caption{Assessment of tuning parameters. The results, which are averaged over all scenarios show that the performance is not particularly sensitive to the choice of parameters. The best result in the left figure can be observed for $W_{\text{MOE}} = 600$ and $W_{\text{MUSIC}} = 700$. In the right figure the \textoverline{FDP} and \textoverline{TPP} are computed for fixed window sizes over different values of  $\alpha$. For small values of $\alpha$ the RD criterion tends to estimate a higher model order $P$, which is reflected in larger \textoverline{FDP} and \textoverline{TPP} values.}
    \label{fig:tuning}
    \vspace{-3mm}
\end{figure}

Generally speaking, in this application, the failure of detecting a person has a far worse impact compared to detecting too many persons. The exact trade-off will depend on scenario specific parameters, such as, the available battery time, the size of the area to be searched, among others. In Fig.~\ref{fig:tuning}, we exemplarily show results for a trade-off that is given by $0.3 \cdot \overline{\text{FDP}} + 0.7 \cdot (1 - \overline{\text{TPP}})$, thus emphasizing \textoverline{TPP} over \textoverline{FDP}. The sensitivity of the person detection w.r.t. different window sizes, i.e. $W_{\text{MOE}}$ and $W_{\text{MUSIC}}$, is reported in Fig.~\ref{fig:w}. The figure shows that, as long as the window sizes are chosen sufficiently large (e.g. $>200$ samples), the results stay in a small range; hence, the results do not swing widely for different window sizes. To reduce the computational complexity, the window sizes should be selected as large as possible, therefore, we selected $W_{\text{MOE}} = 600$ and $W_{\text{MUSIC}} = 700$ as the window sizes.

After fixing the window sizes, we evaluated the sensitivity to the parameter $\alpha$, which tunes the sensitivity of the model order selection criterion. For small values of $\alpha$ the criterion tends to estimate larger values for $P$ and vice versa. This behaviour can be observed in Fig.~\ref{fig:alpha}, where the \textoverline{FDP} decreases with increasing $\alpha$. As shown in Fig.~\ref{fig:alpha}, when increasing $\alpha$, both the \textoverline{TPP} and \textoverline{FDP} are decreasing smoothly, indicating that the algorithm is not very sensitive to the exact choice of $\alpha$. Because in our scenario a high \textoverline{TPP} is very important, small values for $\alpha$ should be selected to detect as many persons as possible. At $\alpha = 3$ we could observe a significant drop in the \textoverline{FDP}, hence we choose $\alpha = 3$ to reduce the expected false alarms, which is also in line with our previous experimental experience. Finally, $L_{\text{st}}$ was set to $L_{\text{st}} = 200$ or approximately 18\,s. 

The MUSIC algorithm was computed using a fixed model order of $P=15$, as it seemed to be favourable to compute the MUSIC pseudo spectrum with a higher, fixed order and afterwards select the $\hat{P}$-highest peaks based on the model order selection. The choice of the model order does not influence the accuracy of the estimated breathing frequency, except for the obvious case when $\hat{P}$ is too small so that not all persons will be detected. Specifically adapting the tuning parameters to improve the \textoverline{FDP} and \textoverline{TPP} in through-wall scenarios did not lead to any noteworthy improvements (see Section~\ref{sec:discussion} for a discussion).

\subsection{Localization Results}

\begin{figure}
	\centering
    \includegraphics[width=0.9\columnwidth]{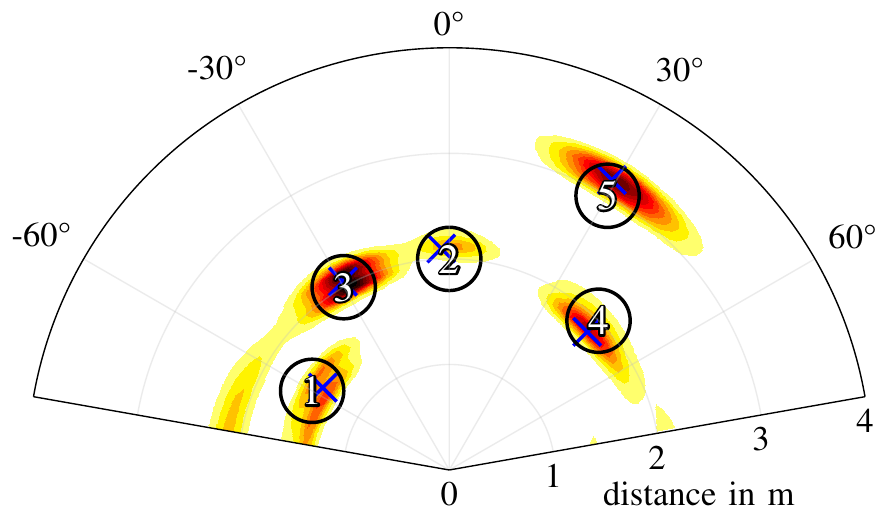}
	\caption{MUSIC spectrum of scenario M16 with free line-of-sight. Circles represent reference person locations as shown in Fig.~\ref{fig:measurement_setup}. Blue crosses mark the estimated locations.}
	\label{fig:musicM16}
    \vspace{-3mm}
\end{figure}
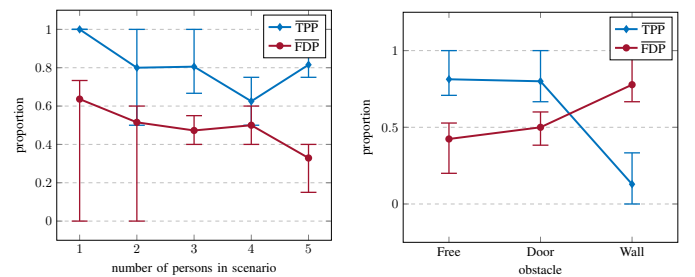
\begin{figure}
    \centering
    \subcaptionbox{\textoverline{TPP} and \textoverline{FDP} over number of persons in scenario. Scenarios with wall are excluded.\label{fig:fdpTPP-person}}[.48\columnwidth]{\resizebox{0.49\columnwidth}{!}{    \begin{tikzpicture}
        \begin{axis}[
        	width = \figurewidth,
	        height = \figureheight,
            xlabel = number of persons in scenario,
            ylabel = proportion,
            ymajorgrids,
            enlarge x limits=0.1,
            enlarge y limits=0.1,
            grid style=dashed,
            cycle list name=matlabcolor,
        ]
        \addplot+[thick,matlabblue,line width=1.5pt,error bars/.cd, error bar style={line width=1pt}, error mark options={rotate=90,matlabblue,mark size=6pt,line width=1pt}, y dir=both, y explicit] table[x=Persons,y=TPP,col sep=comma, y error plus=QuanHiTPP, y error minus=QuanLoTPP] {\PersonResultsTable};
        \addplot+[thick,matlabred,line width=1.5pt,error bars/.cd, error bar style={line width=1pt}, error mark options={rotate=90,matlabred,mark size=6pt,line width=1pt}, y dir=both, y explicit] table[x=Persons,y=FDP,col sep=comma, y error plus=QuanHiFDP, y error minus=QuanLoFDP] {\PersonResultsTable};
        \legend{\textoverline{TPP},\textoverline{FDP}};
        \end{axis}
    \end{tikzpicture}}}
    \hfill
    \subcaptionbox{\textoverline{TPP} and \textoverline{FDP} for different obstacles.\label{fig:fdpTPP-obstacle}}[.48\columnwidth]{\resizebox{0.49\columnwidth}{!}{    \begin{tikzpicture}
        \begin{axis}[
            width = \figurewidth,
	        height = \figureheight,
            xlabel = obstacle,
            ylabel = proportion,
            ymajorgrids,
            ymax=1,
            ymin=0,
            enlarge x limits=0.25,
            enlarge y limits=0.25,
            xtick={0,1,2},
            grid style=dashed,
            xticklabels={Free, Door, Wall},
            cycle list name=matlabcolor,
        ]
        \addplot+[thick,matlabblue,line width=1.5pt,error bars/.cd, error bar style={line width=1pt}, error mark options={rotate=90,matlabblue,mark size=6pt,line width=1pt}, y dir=both, y explicit] table[x=Obstacle,y=TPP,col sep=comma, y error plus=QuanHiTPP, y error minus=QuanLoTPP] {\ObstacleResultsTable};
        \addplot+[thick,matlabred,line width=1.5pt,error bars/.cd, error bar style={line width=1pt}, error mark options={rotate=90,matlabred,mark size=6pt,line width=1pt}, y dir=both, y explicit] table[x=Obstacle,y=FDP,col sep=comma, y error plus=QuanHiFDP, y error minus=QuanLoFDP,] {\ObstacleResultsTable};
        \legend{\textoverline{TPP},\textoverline{FDP}};
        \end{axis}
    \end{tikzpicture}}}
    \caption{\textoverline{TPP} and \textoverline{FDP} averaged for each scenario and grouped by the number of persons present in the left figure and by the kind of obstacle in the right figure. The error bars indicate the 0.25 and 0.75 quantiles.}
    \label{fig:fdpTPP}
    \vspace{-3mm}
\end{figure}

The area of interest is chosen to be $d \in [0, 4.5]$ and $\theta \in [-0.4\pi, 0.4\pi]$. The performance metrics for every measurement are listed in the supplementary material. In most scenarios, our algorithm was able to achieve a localization error of less than 15\,cm, which is less than half of the possible maximum error\footnote{An estimated location was only matched with a reference person if their distance is smaller than 30\,cm, thus the maximum of the location error is bounded to 30\,cm}. Further, it should be noted that the true positions of the persons are not always exact, as in some cases, the test subjects were not sitting precisely at the correct spot, hence, introducing a small uncertainty on the reference location. Fig.~\ref{fig:musicM16} shows an exemplary localization result for scenario M16 with free line-of-sight. The circles represent the true locations with a radius of 30\,cm as shown in Figs.~\ref{fig:measurement_setup} and \ref{fig:field_of_view}. The heat map represents the MUSIC spectrum and the blue crosses represent the estimated locations. In this scenario, all five persons were detected accurately.

As shown in Fig.~\ref{fig:fdpTPP-person}, in free space and door scenarios, there is not a single measurement, in which the algorithm failed to detect at least one person. Hence, in an emergency scenario the location or room would have been definitely searched and with a high probability the other persons would have been also detected. Fig.~\ref{fig:fdpTPP-obstacle} presents the \textoverline{TPP} and \textoverline{FDP} for different obstacles. As expected, the best results are achieved for free space scenarios, followed by scenarios with the door with a slightly worse performance. For wall scenarios the algorithm almost completely breaks down and the results are not satisfactory. This behaviour is not surprising, as the algorithm does not include steps to mitigate the influence of the wall, which is part of our future work. In practice, it would, hence,  be advisable to use the camera information to steer the robot to a door, in case this is possible. A detailed discussion on practical challenges and on the wall scenarios is given in Section~\ref{sec:discussion}.

\begin{figure}
    \centering
    \subcaptionbox{Location error over different number of persons. Scenarios with wall are excluded. \label{fig:locError-person}}[.48\columnwidth]{\resizebox{0.49\columnwidth}{!}{ \begin{tikzpicture}
        \begin{axis}[
        	width = \figurewidth,
	        height = \figureheight,
            xlabel = number of persons in scenario,
            ylabel = median euclidean location error,
            ymajorgrids,
            ymin=0,
            ymax=0.3,
            enlarge x limits=0.1,
            enlarge y limits=0.1,
            grid style=dashed,
            cycle list name=matlabcolor,
        ]
        \addplot+[thick,matlabblue,line width=1.5pt,error bars/.cd, error bar style={line width=1pt}, error mark options={rotate=90,matlabblue,mark size=6pt,line width=1pt}, y dir=both, y explicit] table[x=Persons,y=MedLocError,col sep=comma, y error plus=QuanHiLocError, y error minus=QuanLoLocError] {\PersonResultsTable};
        \end{axis}
    \end{tikzpicture}}}
    \hfill
    \subcaptionbox{Location error for different obstacles \label{fig:locError-obstacle}}[.48\columnwidth]{\resizebox{0.49\columnwidth}{!}{    \begin{tikzpicture}
        \begin{axis}[
            width = \figurewidth,
	        height = \figureheight,
            xlabel = obstacle,
            ylabel = median euclidean location error,
            ymajorgrids,
            ymax=0.3,
            ymin=0,
            enlarge x limits=0.25,
            enlarge y limits=0.25,
            xtick={0,1,2},
            grid style=dashed,
            xticklabels={Free, Door, Wall},
            cycle list name=matlabcolor,
        ]
        \addplot+[thick,matlabblue,line width=1.5pt, error bars/.cd, error bar style={line width=1pt}, error mark options={rotate=90,matlabblue,mark size=6pt,line width=1pt}, y dir=both, y explicit] table[x=Obstacle,y=MedLocError,col sep=comma, y error plus=QuanHiLocError, y error minus=QuanLoLocError] {\ObstacleResultsTable};
        \end{axis}
    \end{tikzpicture}}}
    \caption{Median of the mean Euclidean location error for each scenario and grouped by the number of persons present in the left figure and by the kind of obstacle in the right figure. The error bars indicate the 0.25 and 0.75 quantiles.}
    \label{fig:locError}
    \vspace{-3mm}
\end{figure}
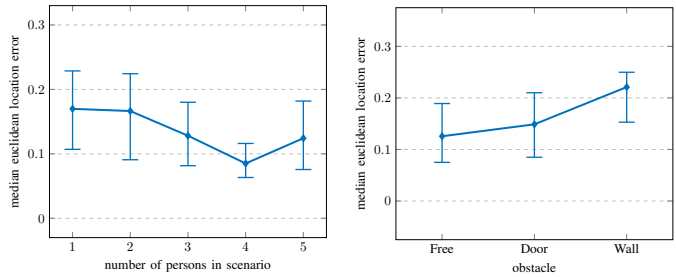

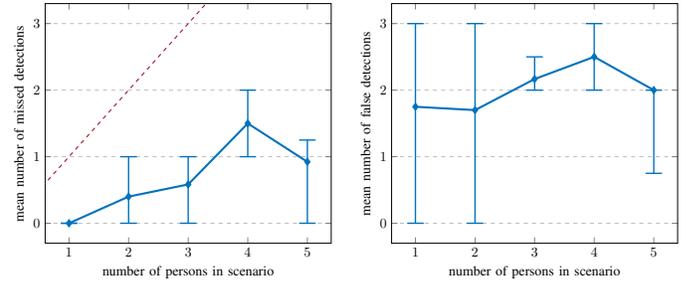
\begin{figure}
    \centering
    \subcaptionbox{Number of missed detections over number of persons. \label{fig:md-person}}[.48\columnwidth]{\resizebox{0.49\columnwidth}{!}{    \begin{tikzpicture}
        \begin{axis}[
        	width = \figurewidth,
	        height = \figureheight,
            ymax=3,
            xlabel = number of persons in scenario,
            ylabel = mean number of missed detections,
            ymajorgrids,
            enlarge x limits=0.1,
            enlarge y limits=0.1,
            grid style=dashed,
            cycle list name=matlabcolor,
        ]
        \addplot+[thick,matlabblue,line width=1.5pt,error bars/.cd, error bar style={line width=1pt}, error mark options={rotate=90,matlabblue,mark size=6pt,line width=1pt}, y dir=both, y explicit] table[x=Persons,y=MD,col sep=comma, y error plus=QuanHiMD, y error minus=QuanLoMD] {\PersonResultsTable};
        \draw[color=matlabred,thick,dashed,domain=0:5,samples=5,smooth] plot(\x,{1*\x});
        \end{axis}
    \end{tikzpicture}}}
    \hfill
    \subcaptionbox{Number of false detections over number of persons. \label{fig:fd-person}}[.48\columnwidth]{\resizebox{0.49\columnwidth}{!}{    \begin{tikzpicture}
        \begin{axis}[
        	width = \figurewidth,
	        height = \figureheight,
            xlabel = number of persons in scenario,
            ylabel = mean number of false detections,
            ymajorgrids,
            ymax=3,
            enlarge x limits=0.1,
            enlarge y limits=0.1,
            grid style=dashed,
            cycle list name=matlabcolor,
        ]
        \addplot+[thick,matlabblue,line width=1.5pt,error bars/.cd, error bar style={line width=1pt}, error mark options={rotate=90,matlabblue,mark size=6pt,line width=1pt}, y dir=both, y explicit] table[x=Persons,y=FD,col sep=comma, y error plus=QuanHiFD, y error minus=QuanLoFD] {\PersonResultsTable};
        \end{axis}
    \end{tikzpicture}}}
    \caption{Mean number of missed and false detections grouped by the number of persons present for all scenarios, excluding scenarios with wall. The error bars indicate the 0.25 and 0.75 quantiles. The dashed red line indicates the maximum possible number of possible missed detections.}
    \label{fig:md-fd}
    \vspace{-3mm}
\end{figure}

In Fig.~\ref{fig:locError}, the localization error is shown for a different number of persons and for different obstacles. Clearly, for scenarios with the wall, the error is larger than in scenarios of free space. Fig.~\ref{fig:md-person} shows the absolute number of missed detections for a different number of persons, the red line indicates the maximum possible number of missed detections. Interestingly, in all one-person-scenarios, the target person was always found. In scenarios with more persons, the number of missed detections does increase slightly, but there is no scenario in which no person was found. The absolute number of false detections, as shown in Fig.~\ref{fig:fd-person}, stays almost constant over the different scenarios.

\subsection{Vital Signs of Detected Persons}

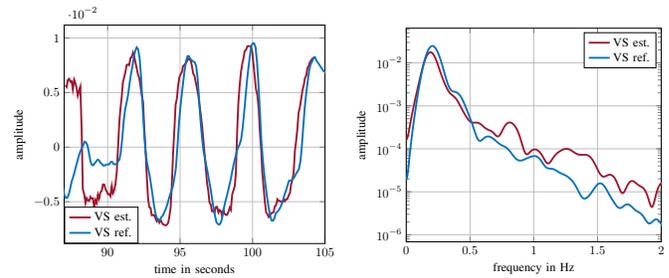
\begin{figure}
    \centering
    \subcaptionbox{Extracted raw breathing signal \label{fig:m16_VS_time_P2}}[.49\columnwidth]{\resizebox{0.5\columnwidth}{!}{\begin{tikzpicture}
        \begin{axis}[
        	width = \figurewidth,
	        height = \figureheight,
            xlabel = time in seconds,
            ylabel = amplitude,
            grid=both,
            xmin=87,
            xmax=105,
            cycle list name=matlabcolor,
            legend style={at={(0.001,0.001)},anchor=south west}
        ]
        \addplot+[thick,matlabred,line width=1.5pt,no marks] table[x=T,y=VS,col sep=comma] {figures/tikz/VS/M16_P2_VS_Radar.csv};
        \addplot+[thick,matlabblue,line width=1.5pt,no marks] table[x expr=\thisrow{T}, y expr=\thisrow{VS}*0.00005, col sep=comma] {figures/tikz/VS/M16_P2_VS_Ref.csv};
        \legend{VS est., VS ref.};
        \end{axis}
    \end{tikzpicture}}}
    \hfill
    \subcaptionbox{Estimated spectrum of P3 \label{fig:m16_VS_freq_P2}}[.49\columnwidth]{\resizebox{0.48\columnwidth}{!}{\begin{tikzpicture}
        \begin{semilogyaxis}[
            width = \figurewidth,
	        height = \figureheight,
            xlabel = frequency in Hz,
            ylabel = amplitude,
            xmax = 2,
            xmin = 0,
            grid,
        ]
        \addplot+[thick,matlabred, line width=1.5pt, mark=none] table[x=F,y expr=\thisrow{VS}*1,col sep=comma] {figures/tikz/VS/M16_P2_Spec_Radar.csv};
        \addplot+[thick,matlabblue, line width=1.5pt, mark=none] table[x=F,y expr=\thisrow{VS}*0.00000001,col sep=comma] {figures/tikz/VS/M16_P2_Spec_Ref.csv};
        \legend{VS est., VS ref.};
    \end{semilogyaxis}
\end{tikzpicture}}}
    \caption{In the left figure, the extracted raw breathing signal of Person 3 from M16 is compared with the reference signal. In the right figure, the resulting estimated spectrum is shown.}
    \label{fig:m16_VS_freq_P2_all}
    \vspace{-3mm}
\end{figure}

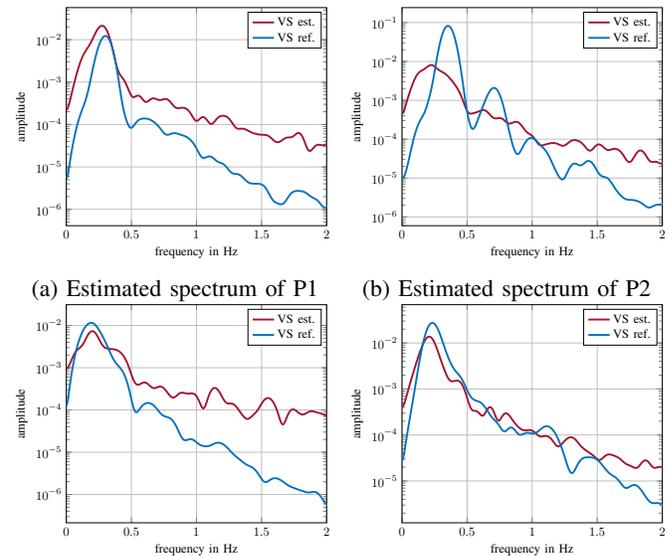
\begin{figure}
    \centering
    \subcaptionbox{Estimated spectrum of P1\label{fig:m16_VS_freq_P1}}[.49\columnwidth]{\resizebox{0.49\columnwidth}{!}{\begin{tikzpicture}
        \begin{semilogyaxis}[
            width = \figurewidth,
	        height = \figureheight,
            xlabel = frequency in Hz,
            ylabel = amplitude,
            xmax = 2,
            xmin = 0,
            grid,
        ]
        \addplot+[thick,matlabred, line width=1.5pt, mark=none] table[x=F,y expr=\thisrow{VS}*1,col sep=comma] {figures/tikz/VS/M16_P1_Spec_Radar.csv};
        \addplot+[thick,matlabblue, line width=1.5pt, mark=none] table[x=F,y expr=\thisrow{VS}*0.00000001,col sep=comma] {figures/tikz/VS/M16_P1_Spec_Ref.csv};
        \legend{VS est., VS ref.};
        \end{semilogyaxis}
    \end{tikzpicture}}}
    \subcaptionbox{Estimated spectrum of P2 \label{fig:m16_VS_freq_P3}}[.49\columnwidth]{\resizebox{0.49\columnwidth}{!}{\begin{tikzpicture}
        \begin{semilogyaxis}[
            width = \figurewidth,
	        height = \figureheight,
            xlabel = frequency in Hz,
            ylabel = amplitude,
            xmax = 2,
            xmin = 0,
            grid,
        ]
        \addplot+[thick,matlabred, line width=1.5pt, mark=none] table[x=F,y expr=\thisrow{VS}*1,col sep=comma] {figures/tikz/VS/M16_P3_Spec_Radar.csv};
        \addplot+[thick,matlabblue, line width=1.5pt, mark=none] table[x=F,y expr=\thisrow{VS}*0.00000001,col sep=comma] {figures/tikz/VS/M16_P3_Spec_Ref.csv};
        \legend{VS est., VS ref.};
        \end{semilogyaxis}
    \end{tikzpicture}}}\\
    \subcaptionbox{Estimated spectrum of P4\label{fig:m16_VS_freq_P4}}[.49\columnwidth]{\resizebox{0.49\columnwidth}{!}{\begin{tikzpicture}
        \begin{semilogyaxis}[
            width = \figurewidth,
	        height = \figureheight,
            xlabel = frequency in Hz,
            ylabel = amplitude,
            xmax = 2,
            xmin = 0,
            grid,
        ]
        \addplot+[thick,matlabred, line width=1.5pt, mark=none] table[x=F,y expr=\thisrow{VS}*1,col sep=comma] {figures/tikz/VS/M16_P4_Spec_Radar.csv};
        \addplot+[thick,matlabblue, line width=1.5pt, mark=none] table[x=F,y expr=\thisrow{VS}*0.00000001,col sep=comma] {figures/tikz/VS/M16_P4_Spec_Ref.csv};
        \legend{VS est., VS ref.};
        \end{semilogyaxis}
    \end{tikzpicture}}}
    \subcaptionbox{Estimated spectrum of P5 \label{fig:m16_VS_freq_P5}}[.49\columnwidth]{\resizebox{0.49\columnwidth}{!}{\begin{tikzpicture}
        \begin{semilogyaxis}[
            width = \figurewidth,
	        height = \figureheight,
            xlabel = frequency in Hz,
            ylabel = amplitude,
            xmax = 2,
            xmin = 0,
            grid,
        ]
        \addplot+[thick,matlabred, line width=1.5pt, mark=none] table[x=F,y expr=\thisrow{VS}*1,col sep=comma] {figures/tikz/VS/M16_P5_Spec_Radar.csv};
        \addplot+[thick,matlabblue, line width=1.5pt, mark=none] table[x=F,y expr=\thisrow{VS}*0.00000001,col sep=comma] {figures/tikz/VS/M16_P5_Spec_Ref.csv};
        \legend{VS est., VS ref.};
        \end{semilogyaxis}
    \end{tikzpicture}}}
    \caption{Estimated breathing frequencies of persons from M16 compared to the reference data. The depicted spectra are averaged over all time windows as shown in Fig.~\ref{fig:m16_vs_over_time}.}
    \label{fig:m16_VS_freq}
    \vspace{-3mm}
\end{figure}

\begin{figure}
    \centering
    \resizebox{0.8\columnwidth}{!}{
    \begin{tikzpicture}

    \begin{groupplot}[
        scale = 1.75,
        group style={group name=my plots,
                    group size=1 by 5,
                    xlabels at=edge bottom,
                    xticklabels at=edge bottom,
                    ylabels at=edge left,
                    yticklabels at=edge left,
                    vertical sep=10pt},
        width=\figurewidth,
        height=2.75cm,
        xlabel=segment,
        ymin=0,
        ymax=0.5,
        xmin = 1,
        xmax = 10,
        xtick={1,...,10},
        grid,
        cycle list name=matlabcolor,
	    legend pos=north east,
        ]
        \nextgroupplot
        \addplot+[matlabred,line width=1.5pt, mark=none] table[x index=0, y index=6,col sep=comma] {figures/resultsPlots/estVSrefBreath-ID16.csv}; 
        \addplot+[matlabblue,line width=1.5pt, mark=none] table[x index=0, y index=1,col sep=comma] {figures/resultsPlots/estVSrefBreath-ID16.csv}; 
        \legend{P1 est., P1 ref.};
        \nextgroupplot[legend pos=south east]
        \addplot+[matlabred,line width=1.5pt, unbounded coords=jump, mark=none] table[x index=0, y index=8,col sep=comma] {figures/resultsPlots/estVSrefBreath-ID16.csv}; 
        \addplot+[matlabblue,line width=1.5pt, mark=none] table[x index=0, y index=3,col sep=comma] {figures/resultsPlots/estVSrefBreath-ID16.csv}; 
        \legend{P2 est., P2 ref.};
        \nextgroupplot[ylabel = frequency in Hz]
        \addplot+[matlabred,line width=1.5pt, unbounded coords=jump, mark=none] table[x index=0, y index=7,col sep=comma] {figures/resultsPlots/estVSrefBreath-ID16.csv}; 
        \addplot+[matlabblue,line width=1.5pt, mark=none] table[x index=0, y index=2,col sep=comma] {figures/resultsPlots/estVSrefBreath-ID16.csv}; 
        \legend{P3 est., P3 ref.};
        \nextgroupplot
        \addplot+[matlabred,line width=1.5pt, mark=none] table[x index=0, y index=9,col sep=comma] {figures/resultsPlots/estVSrefBreath-ID16.csv}; 
        \addplot+[matlabblue,line width=1.5pt, mark=none] table[x index=0, y index=4,col sep=comma] {figures/resultsPlots/estVSrefBreath-ID16.csv}; 
        \legend{P4 est., P4 ref.};
        \nextgroupplot
        \addplot+[matlabred,line width=1.5pt, mark=none] table[x index=0, y index=10,col sep=comma] {figures/resultsPlots/estVSrefBreath-ID16.csv}; 
        \addplot+[matlabblue,line width=1.5pt, mark=none] table[x index=0, y index=5,col sep=comma] {figures/resultsPlots/estVSrefBreath-ID16.csv}; 
        \legend{P5 est., P5 ref.};
    \end{groupplot}
\end{tikzpicture}}
    \caption{Comparison of estimated breathing frequencies with reference breathing frequencies over different time windows. The missing data point for Person 3 indicates that this person was not detected in this window.}
    \label{fig:m16_vs_over_time}
    \vspace{-3mm}
\end{figure}
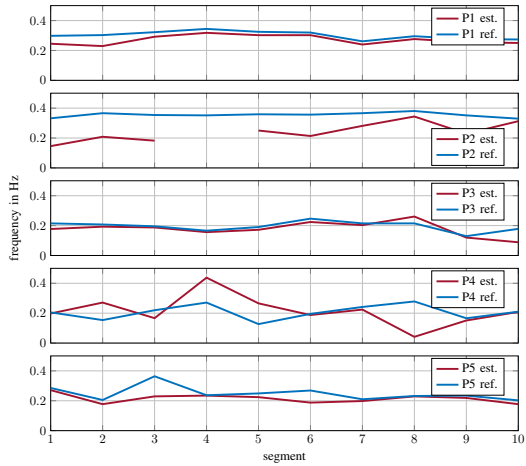

In Figs.~\ref{fig:m16_VS_freq_P2_all} and \ref{fig:m16_VS_freq}, the extracted breathing signals for M16 are shown for the estimated locations from Fig.~\ref{fig:musicM16}. In Fig.~\ref{fig:m16_VS_time_P2}, we display an exemplary raw extracted breathing signal, directly compared with the reference signal. The estimated signal has an amplitude in meter and the reference signal is scaled accordingly. For this segment, both signals are a very close fit and, at the beginning of the segment, the radar-based vital signs seem to even provide a better estimate than the reference signal. The depicted periodograms are averaged over all segments. As it can be seen, both for the reference and the radar-based estimate, the breathing frequency estimate that uses the strongest peak of the periodogram is at about 0.3\,Hz. 

Further examples of breathing frequency estimates are shown in Fig.~\ref{fig:m16_vs_over_time}. For Persons 1, 3 and 5, one can observe a close fit of the reference and estimated breathing frequencies, even small fluctuations over the different segments are correctly displayed. The estimate for Person 2 has a constant offset to the reference frequency. This could be explained by the fact that Person 2 was difficult to localize, as it was not detected in Segment 4. For Person 4, the estimated and reference signals are not always aligned, however, as shown in Fig.~\ref{fig:m16_VS_freq_P4}, the overall respiration frequency estimate is still accurate. 

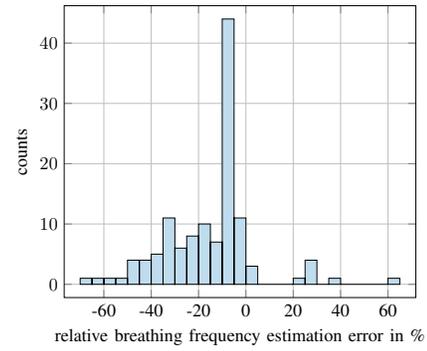
\begin{figure}
    \centering
    \resizebox{0.65\columnwidth}{!}{
    \begin{tikzpicture}
\begin{axis}[
    enlargelimits=0.05,
    ylabel={counts},
    xlabel={relative breathing frequency estimation error in \%},
    xtick={-0.6,-0.4,-0.2,0,0.2,0.4,0.6},
    xticklabels={-60,-40,-20,0,20,40,60},
    grid=both,
]
\addplot[ybar interval,mark=none,fill=matlabblue!25] table [x=Coordinate, y=Frequency, col sep=comma] {figures/tikz/RespFreqErrorRel_Hist.csv};
\end{axis}
\end{tikzpicture}}
    \caption{Relative breathing frequency estimation error for all experiments and all estimated persons.}
    \label{fig:histogram}
    \vspace{-3mm}
\end{figure}

An overview of all measured vital signs is given in Fig.~\ref{fig:histogram}, where the relative error in the estimation of all detected persons' breathing frequencies is shown. In most cases, the breathing frequency is estimated reasonably close to the reference breathing frequency with an error of less than 10\%. Nevertheless, the histogram also shows a significant number of errors greater than 30\%, which must be considered as a large deviation from the reference signal. 

When taking a close look at Fig.~\ref{fig:histogram}, it can be noticed that the breathing frequencies, on average, tend to be underestimated. A discussion on some possible reasons for this result is given in what follows: Firstly, clutter could shift the estimated frequencies lower, as we observed high amplitudes for low-frequency components in empty-room scenarios. When the clutter is not removed completely, or erroneous locations are estimated, we assume that this could lead to an overall negative bias of the estimated frequencies. Secondly, the bias towards lower frequencies could be a systematic error where a person who is breathing slower, which usually also implies deeper, will be detected more prominently since the amplitude of the signal is higher. When such a person is sitting side-by-side with a shallow (and quickly) breathing person, the signal of the deep breathing person might interfere with the shallow breathing signal. Thirdly, the bias towards lower frequencies could also be due to the fact that a person is not sitting completely still during a measurement, but does some very slow body movements (such as rocking forward and backward), which could be mixed up with the chest movement from the breathing and would also lead to a lower estimated breathing frequency.

\begin{figure}
    \centering
    \subcaptionbox{Extracted raw breathing signal of Person 2. \label{fig:m8_VS_time_P2}}[.49\columnwidth]{\resizebox{0.49\columnwidth}{!}{\begin{tikzpicture}
        \begin{axis}[
        	width = \figurewidth,
	        height = \figureheight,
            xlabel = time in seconds,
            ylabel = amplitude,
            grid=both,
            xmin=87,
            xmax=105,
            cycle list name=matlabcolor,
        ]
        \addplot+[thick,matlabred,line width=1.5pt,no marks] table[x=T,y=VS,col sep=comma,] {figures/tikz/VS/M8_P2_VS_Radar.csv};
        \addplot+[thick,matlabblue,line width=1.5pt,no marks] table[x expr=\thisrow{T}, y expr=\thisrow{VS}*0.0001, col sep=comma,] {figures/tikz/VS/M8_P2_VS_Ref.csv};
        \legend{VS est., VS ref.};
        \end{axis}
    \end{tikzpicture}}}
    \hfill
    \subcaptionbox{Spectral estimate of extracted breathing signal. \label{fig:m8_VS_freq_P2}}[.49\columnwidth]{\resizebox{0.49\columnwidth}{!}{\begin{tikzpicture}
        \begin{semilogyaxis}[
            width = \figurewidth,
	        height = \figureheight,
            xlabel = frequency in Hz,
            ylabel = amplitude,
            xmax = 2,
            xmin = 0,
            grid,
        ]
        \addplot+[thick,matlabred, line width=1.5pt, mark=none] table[x=F,y expr=\thisrow{VS}*1,col sep=comma] {figures/tikz/VS/M8_P2_Spec_Radar.csv};
        \addplot+[thick,matlabblue, line width=1.5pt, mark=none] table[x=F,y expr=\thisrow{VS}*0.00000001,col sep=comma] {figures/tikz/VS/M8_P2_Spec_Ref.csv};
        \legend{VS est., VS ref.};
        \end{semilogyaxis}
    \end{tikzpicture}}}
    \caption{Vital sign estimation in M8, where the line of sight between radar and person is obstructed by a door.}
    \label{fig:m8_VS_freq_P2_all}
    \vspace{-3mm}
\end{figure}
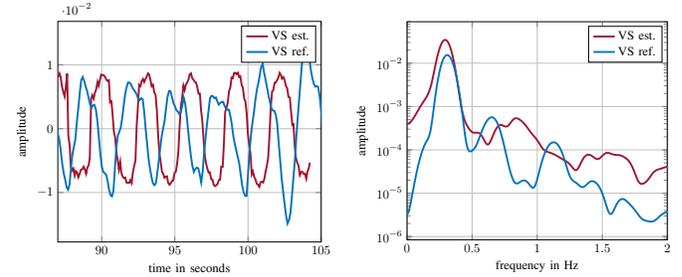

Fig.~\ref{fig:m8_VS_freq_P2_all} depicts the extracted vital signs of a person for a measurement, where the line of sight between person and radar is obstructed by a wooden door. The estimated breathing signal in Fig.~\ref{fig:m8_VS_freq_P2} has a time offset compared to the reference signal. The reason for this could be an additional signal delay due to the door. Nevertheless, in Fig.~\ref{fig:m8_VS_freq_P2} it can be seen that the spectral estimate of the radar vital sign signal indicates a peak that aligns with that of the reference signal. 

\section{Discussion, Limitations \& Future Work}
\label{sec:discussion}

The overall results of the presented method are encouraging, but it should be emphasized that the proposed method serves as a baseline approach for the detection and estimation of vital sign parameters with a radar system. In what follows, we discuss practical challenges, some potential limitations and possible future improvements.

We begin the discussion by commenting on practical challenges. Multiple different types of practical challenges can be encountered some of which are related to the surrounding environment, while others are related to the used sensing platform and signal processing framework and some are systematic to the problem.
    
\begin{figure}
	\centering
    \includegraphics[width=0.9\columnwidth]{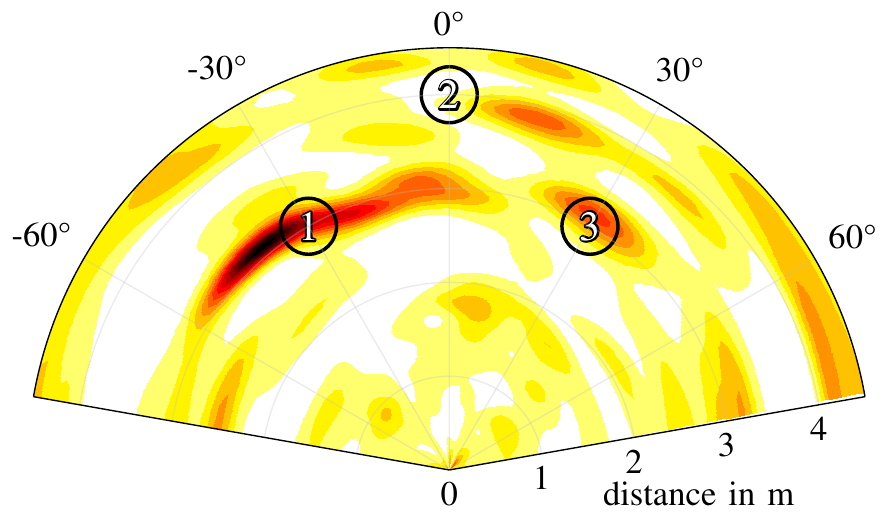}
	\caption{MUSIC spectrum of measurement M27-1. Circles represent reference person locations.}
	\label{fig:musicM271}
    \vspace{-3mm}
\end{figure}

\emph{Environmental challenges} include an unknown and dangerous terrain, in which the robot has to navigate, or unknown harsh local conditions like severe smoke, dust, obstacles, fire, or radiation which could influence the robot navigation and could damage the physical hardware of the robot or radar. A further emergency response specific challenge is that the material and composition of obstacles, such as walls or doors are generally unknown, which could lead to further challenges in the signal processing chain. Challenges concerning through-wall scenarios are discussed, e.g., in \cite{Amin.2011}. The attenuation caused by the wall will lead to inaccuracies in the measurement of the signal, due to noise and the limited dynamic range of the radar. This effect is increased as the signal has to travel trough the wall twice. Fig.~\ref{fig:musicM271} shows an exemplary through-wall MUSIC pseudo spectrum, where a decreased SNR can be observed. Further, the used cinder blocks contain hollow chambers. These periodic repetitions of the hollow chambers inside of the cinder blocks can lead to additional signal distortions \cite[p. 292]{Amin.2011}. The used emergenCITY Scout robot has already demonstrated its practical capabilities in the Zwentendorf nuclear power plant in Austria to navigate, create a radiation map, and to provide information on injured people, or find hidden sources of radiation. Overcoming further environmental challenges and making the signal processing chain more resilient will likely require new approaches that are benchmarked on extensive measurement campaigns, which include a variety of the above mentioned aspects.

\emph{Practical challenges regarding the sensing platform and signal processing} include (i) the limited battery time of the robot, (ii) the limited transmit power of the used radar and (iii) the size of the antenna array. To increase the battery time of the robot, the efficiency of the algorithms has to be increased, or new dedicated algorithms of lower computational complexity have to be developed. A practical way to trade-off battery power against localization accuracy within the proposed framework, is to adapt the size of the grid search in the 2D-MUSIC algorithm. In practice, in future, this grid-size parameter may be adapted to the given situation. A higher accuracy may be obtained by increasing the transmit power of the radar and/or increasing the number of transmit/receive antennas. This would help in clutter removal and signal reconstruction, but could also reduce the battery time.

\emph{Systematic practical challenges} inherent to the problem include (i) the time-delay until a breathing frequency can be reliably estimated, (ii) false discoveries and (iii) low received signal amplitudes, especially in trough-wall-scenarios. With typical breathing frequencies in the range of 0.2\,Hz to 0.5\,Hz, at least a few seconds of data are required to reliably detect and measure a vital signal. This hinders a fast and efficient exploration of the environment; hence, further research in sensing while moving needs to be considered. A fast exploration is also delayed by false discoveries which steer valuable resources in the wrong direction. Here an adaptation of FDR control methods, such as \cite{Machkour.2021} could be beneficial. Concerning the received amplitudes, a typical round-trip of the signal has a loss of over 40\,dB in solid concrete and around 2\,dB in plywood at 3\,GHz \cite[p.38]{Amin.2011}. Hence, determining the material of the wall and adapting the radar signal and amplitude accordingly could improve the results. However, this would require access to the radar signal generation, which is not possible with the current device. The wall material could be estimated by making use of available prior knowledge, cameras or infrared sensors that are also mounted on the emergenCITY Scout robot.

Currently, the presented framework is limited to stationary sensing. Overcoming this limitation has some advantages, e.g., more survivors may be detected because the robot could explore while sensing, or the radar aperture could be extended by deriving a synthetic aperture approach. However, sensing while moving leads to  challenges in (i) clutter removal, (ii) motion compensation and (iii) in determining the relative location of the robot. In the proposed  method, the SMA filter is used to remove the clutter from the received signal. This is possible because the clutter is stationary, whereas the target person is moving (breathing). In contrast, removing the clutter while moving would imply the need for more advanced clutter removal methods. Furthermore, small vibrations induced by the robot while moving, could be misinterpreted as vitals signs because the amplitude of the vital signs and the vibrations are in the same amplitude range. This leads to the necessity to compensate for these small vibrations, e.g., by measuring them on the robot. To extract the vital signs, the exact location of a target in relation to the robot has to be known and continuously updated during the movement of the robot. A very accurate update of the location (more accurate than GPS which is also not always available in emergency response settings) of the robot would thus be needed to continuously track the movement of the robot in relation to the target.

In the remainder, possible ideas and suggestions for future work and improvements are given. Firstly, using the estimated vital signs, it might be possible to further decrease false positives by discarding locations without a valid breathing frequency. Secondly, it should be also possible to theoretically estimate the heart rate of a person, but using straightforward approaches, a reliable estimation did not seem possible. However, a more sophisticated signal processing chain may be capable to estimate the heart rate after extracting all breathing-related signal components. Therefore, it might be beneficial to estimate the frequencies based on an empirical mode decomposition \cite{Shyu.2019} or the wavelet transform \cite{He.2020}. Thirdly, the currently used phase unwrapping algorithm introduces errors. Hence, it might be beneficial to use more sophisticated phase unwrapping algorithms as presented in \cite{Wang.2014, Baldi.2002}. Finally, as discussed previously, further work is required for scenarios with a non-homogeneous stone wall. Mitigating the effects of heterogeneous walls would require integrating knowledge of the wall material (or estimated wall parameters) into the algorithms, which may lead to improved through-wall localization results. Additionally, in through-wall scenarios the current model order estimation algorithm breaks down, and hence, persons were not reliably detected. Introducing a robust model order estimation algorithm, e.g., a robust Bayesian Information Criterion \cite{Schroth.2021}, could potentially lead to a significantly increased performance.

Summarizing, we showed that the presented baseline algorithm is a valuable proof of concept. We will and will encourage others to further develop the proposed algorithm. This will improve upon existing approaches for radar-based multi-person detection, localization and vital sign estimation with different types of obstacles.
\section{Conclusion}
\label{sec:conclusion}
A comprehensive signal processing chain for radar-based multi-person detection, localization and breathing frequency estimation was proposed and its performance was evaluated based on a new challenging emergency response open accessible benchmark dataset. The application of the radar-based person localization and vital sign estimation on a mobile rescue robot used for autonomously exploring a disaster site opens up new perspectives for disaster response applications as well as for research in related areas, such as, digital healthcare or assisted living. The collected dataset includes a large variety of different scenarios which enables the development and benchmarking of future algorithms. We hope to see other researchers using the publicly available codes and data and further improve upon our results so as to progress in this important application. 

\section*{Acknowledgment}

The authors wish to thank the volunteer students who participated in the data collection.

\printbibliography

@article{Acar.2021,
 author = {Acar, Yunus Emre and Saritas, Ismail and Yaldiz, Ercan},
 year = {2021},
 title = {{An experimental study: Detecting the respiration rates of multiple stationary human targets by stepped frequency continuous wave radar}},
 pages = {108268},
 volume = {167},
 journal = {{Measurement}}
}

@book{Amin.2011,
 year = {2011},
 title = {{Through-the-Wall Radar Imaging}},
 address = {Boca Raton, Florida},
 publisher = {{CRC Press}},
 isbn = {9781439814765},
 editor = {Amin, Moeness G.}
}

@article{Amin.2016,
 author = {Amin, Moeness G. and Zhang, Yimin D. and Ahmad, Fauzia and Ho, K. DominicC.},
 year = {2016},
 title = {{Radar Signal Processing for Elderly Fall Detection: The future for in-home monitoring}},
 pages = {71--80},
 volume = {33},
 number = {2},
 journal = {{IEEE Signal Processing Magazine}}
}

@article{Baldi.2002,
 author = {Baldi, A. and Bertolino, F. and Ginesu, F.},
 year = {2002},
 title = {{On the performance of some unwrapping algorithms}},
 pages = {313--330},
 volume = {37},
 number = {4},
 journal = {{Optics and Lasers in Engineering}}
}

@inproceedings{Belfiori.2012,
 author = {Belfiori, Francesco and {van Rossum}, Wim and Hoogeboom, Peter},
 title = {{Application of 2D MUSIC algorithm to range-azimuth FMCW radar data}},
 url = {https://ieeexplore.ieee.org/document/6450737},
 pages = {242--245},
 publisher = {IEEE},
 isbn = {978-1-4673-2471-7},
 booktitle = {{2012 9th European Radar Conference (EuRAD 2012)}},
 year = {2012}
}

@inproceedings{Daun.2021,
 author = {Daun, Kevin and Schnaubelt, Marius and Kohlbrecher, Stefan and von Stryk, Oskar},
 title = {{HectorGrapher: Continuous-time Lidar SLAM with Multi-resolution Signed Distance Function Registration for Challenging Terrain}},
 pages = {152--159},
 publisher = {IEEE},
 isbn = {978-1-6654-1764-8},
 booktitle = {{2021 IEEE International Symposium on Safety, Security, and Rescue Robotics (SSRR)}},
 year = {2021}
}

@article{Debes.2009,
 author = {Debes, C. and Amin, Moeness G. and Zoubir, Abdelhak M.},
 year = {2009},
 title = {{Target Detection in Single- and Multiple-View Through-the-Wall Radar Imaging}},
 pages = {1349--1361},
 volume = {47},
 number = {5},
 journal = {{IEEE Transactions on Geoscience and Remote Sensing}}
}

@article{Debes.2011,
 author = {Debes, Christian and Hahn, J{\"u}rgen and Zoubir, Abdelhak M. and Amin, Moeness G.},
 year = {2011},
 title = {{Target Discrimination and Classification in Through-the-Wall Radar Imaging}},
 pages = {4664--4676},
 volume = {59},
 number = {10},
 journal = {{IEEE Transactions on Signal Processing}}
}

@article{Feger.2009,
 author = {Feger, R. and Wagner, C. and Schuster, S. and Scheiblhofer, S. and Jager, H. and Stelzer, A.},
 year = {2009},
 title = {{A 77-GHz FMCW MIMO Radar Based on an SiGe Single-Chip Transceiver}},
 pages = {1020--1035},
 volume = {57},
 number = {5},
 journal = {{IEEE Transactions on Microwave Theory and Techniques}}
}

@incollection{Fioranelli.2023,
 author = {Fioranelli, Francesco and Guendel, Ronny G. and Kruse, Nicolas C. and Yarovoy, Alexander},
 title = {{Radar Sensing in Healthcare: Challenges and Achievements in Human Activity Classification {\&} Vital Signs Monitoring}},
 pages = {492--504},
 volume = {13920},
 publisher = {{Springer Nature Switzerland}},
 isbn = {978-3-031-34959-1},
 series = {{Lecture Notes in Computer Science}},
 editor = {Rojas, Ignacio and Valenzuela, Olga and {Rojas Ruiz}, Fernando and Herrera, Luis Javier and Ortu{\~n}o, Francisco},
 booktitle = {{Bioinformatics and Biomedical Engineering}},
 year = {2023}
}

@book{Garello.2008,
 year = {2008},
 title = {{Two-dimensional signal analysis}},
 address = {London},
 publisher = {ISTE},
 isbn = {978-1-84821-018-9},
 series = {{Digital signal and image processing series}},
 editor = {Garello, Ren{\'e}}
}

@article{Gentilho.2020,
 author = {Gentilho, Edno and Scalassara, Paulo Rogerio and Abr{\~a}o, Taufik},
 year = {2020},
 title = {{Direction-of-Arrival Estimation Methods: A Performance-Complexity Tradeoff Perspective}},
 pages = {239--256},
 volume = {92},
 number = {2},
 journal = {{Journal of Signal Processing Systems}}
}

@article{Gupta.2022,
 author = {Gupta, Khushi and {M. B.}, Srinivas and J, Soumya and Pandey, Om Jee and Cenkeramaddi, Linga Reddy},
 year = {2022},
 title = {{Automatic Contact-Less Monitoring of Breathing Rate and Heart Rate Utilizing the Fusion of mmWave Radar and Camera Steering System}},
 pages = {22179--22191},
 volume = {22},
 number = {22},
 journal = {{IEEE Sensors Journal}}
}

@article{Han.2022,
 author = {Han, Weiqiao and Dai, Shaozhang and Yuce, Mehmet Rasit},
 year = {2022},
 title = {{Real-Time Contactless Respiration Monitoring From a Radar Sensor Using Image Processing Method}},
 pages = {19020--19029},
 volume = {22},
 number = {19},
 journal = {{IEEE Sensors Journal}}
}

@article{He.2020,
 author = {He, Mi and Nian, Yongjian and Xu, Luping and Qiao, Lihong and Wang, Wenwu},
 year = {2020},
 title = {{Adaptive Separation of Respiratory and Heartbeat Signals among Multiple People Based on Empirical Wavelet Transform Using UWB Radar}},
 volume = {20},
 number = {17},
 journal = {{Sensors}}
}

@inproceedings{Hollick.2019,
 author = {Hollick, Matthias and Hofmeister, Anne and Engels, Jens Ivo and Freisleben, Bernd and Hornung, Gerrit and Klein, Anja and Knodt, Mich{\`e}le and Lorenz, Imke and M{\"u}hlh{\"a}user, Max and Pelz, Peter and Rudolph-Cleff, Annette and Steinmetz, Ralf and Steinke, Florian and von Stryk, Oskar},
 title = {{The Emergency Responsive Digital City}},
 url = {https://www.sim.informatik.tu-darmstadt.de/publ/download/2019_emergenCITY_WCRRAM.pdf},
 pages = {136--139},
 booktitle = {{World Congress on Resilience, Reliability and Asset Management 2019}},
 year = {2019}
}

@inproceedings{Huang.2015,
 author = {Huang, Xinming and Sun, Ling and Tian, Tian and Huang, Zeyan and Clancy, Edward},
 title = {{Real-time non-contact infant respiratory monitoring using UWB radar}},
 pages = {493--496},
 publisher = {IEEE},
 isbn = {978-1-4673-7004-2},
 booktitle = {{2015 IEEE 16th International Conference on Communication Technology (ICCT)}},
 year = {2015}
}

@article{Huang.2020,
 author = {Huang, Lidong and Wang, Xianpeng and Huang, Mengxing and Wan, Liangtian and Han, Zhiguang and Yang, Yongqin},
 year = {2020},
 title = {{An Implementation Scheme of Range and Angular Measurements for FMCW MIMO Radar via Sparse Spectrum Fitting}},
 pages = {389},
 volume = {9},
 number = {3},
 journal = {{Electronics}}
}

@article{Islam.2022,
 author = {Islam, Fabliha Bushra and Lee, Jae-Min and Kim, Dong-Seong},
 year = {2022},
 title = {{Smart factory floor safety monitoring using UWB sensor}},
 pages = {412--425},
 volume = {16},
 number = {7},
 journal = {{IET Science, Measurement {\&} Technology}}
}

@article{Jain.2012,
 author = {Jain, Sachin K. and Singh, S. N.},
 year = {2012},
 title = {{Exact Model Order ESPRIT Technique for Harmonics and Interharmonics Estimation}},
 pages = {1915--1923},
 volume = {61},
 number = {7},
 journal = {{IEEE Transactions on Instrumentation and Measurement}}
}

@article{Jia.2019,
 author = {Jia, Yong and Guo, Yong and Yan, Chao and Sheng, Haoxuan and Cui, Guolong and Zhong, Xiaoling},
 year = {2019},
 title = {{Detection and Localization for Multiple Stationary Human Targets Based on Cross-Correlation of Dual-Station SFCW Radars}},
 pages = {1428},
 volume = {11},
 number = {12},
 journal = {{Remote Sensing}}
}

@article{Kakouche.2021b,
 author = {Kakouche, Ibrahim and Maali, Abdelmadjid and {El Korso}, Mohammed Nabil and Mesloub, Ammar and Azzaz, Mohamed Salah},
 year = {2021},
 title = {{Fast and cost-effective method for non-contact respiration rate tracking using UWB impulse radar}},
 pages = {112814},
 volume = {329},
 journal = {{Sensors and Actuators A: Physical}}
}

@article{Kakouche.2021c,
 author = {Kakouche, Ibrahim and Abadlia, Hamza and {El Korso}, Mohammed Nabil and Mesloub, Ammar and Maali, Abdelmadjid and Azzaz, Mohamed Salah},
 year = {2021},
 title = {{Joint Vital Signs and Position Estimation of Multiple Persons Using SIMO Radar}},
 pages = {2805},
 volume = {10},
 number = {22},
 journal = {{Electronics}}
}

@article{Kakouche.2022,
 author = {Kakouche, Ibrahim and Maali, Abdelmadjid and {El Korso}, Mohammed Nabil and Mesloub, Ammar and Azzaz, Mohamed Salah},
 year = {2022},
 title = {{Non-contact measurement of respiration and heart rates based on subspace methods and iterative notch filter using UWB impulse radar}},
 pages = {035401},
 volume = {55},
 number = {3},
 journal = {{Journal of Physics D: Applied Physics}}
}

@article{Khan.2017,
 author = {Khan, Faheem and Cho, Sung Ho},
 year = {2017},
 title = {{A Detailed Algorithm for Vital Sign Monitoring of a Stationary/Non-Stationary Human through IR-UWB Radar}},
 volume = {17},
 number = {2},
 journal = {{Sensors}}
}

@article{Kim.2015,
 author = {Kim, Jong-Hwan and Starr, Joseph W. and Lattimer, Brian Y.},
 year = {2015},
 title = {{Firefighting Robot Stereo Infrared Vision and Radar Sensor Fusion for Imaging through Smoke}},
 pages = {823--845},
 volume = {51},
 number = {4},
 journal = {{Fire Technology}}
}

@article{Kozlov.2022,
 author = {Kozlov, Roman and Gavrilov, Konstantin and Shevgunov, Timofey and Kirdyashkin, Vladimir},
 year = {2022},
 title = {{Stepped-Frequency Continuous-Wave Signal Processing Method for Human Detection Using Radars for Sensing Rooms through the Wall}},
 pages = {79},
 volume = {7},
 number = {3},
 journal = {{Inventions}}
}

@book{Kraus.1988,
 author = {Kraus, John Daniel},
 year = {1988},
 title = {{Antennas}},
 address = {New York},
 edition = {2. ed.},
 publisher = {McGraw-Hill},
 isbn = {0070354227},
 series = {{McGraw-Hill series in electrical engineering Radar and antennas}}
}

@inproceedings{KruijffKorbayova.2021,
 author = {Kruijff-Korbayova, Ivana and Grafe, Robert and Heidemann, Nils and Berrang, Alexander and Hussung, Cai and Willms, Christian and Fettke, Peter and Beul, Marius and Quenzel, Jan and Schleich, Daniel and Behnke, Sven and Tiemann, Janis and Guldenring, Johannes and Patchou, Manuel and Arendt, Christian and Wietfeld, Christian and Daun, Kevin and Schnaubelt, Marius and von Stryk, Oskar and Lel, Alexander and Miller, Alexander and Rohrig, Christof and Strassmann, Thomas and Barz, Thomas and Soltau, Stefan and Kremer, Felix and Rilling, Stefan and Haseloff, Rohan and Grobelny, Stefan and Leinweber, Artur and Senkowski, Gerhard and Thurow, Marc and Slomma, Dominik and Surmann, Hartmut},
 title = {{German Rescue Robotics Center (DRZ): A Holistic Approach for Robotic Systems Assisting in Emergency Response}},
 pages = {138--145},
 publisher = {IEEE},
 isbn = {978-1-6654-1764-8},
 booktitle = {{2021 IEEE International Symposium on Safety, Security, and Rescue Robotics (SSRR)}},
 year = {2021}
}

@inproceedings{Kusmadi.2015,
 author = {Kusmadi and Munir, Achmad},
 title = {{Simulation design of compact stepped-frequency continuous-wave through-wall radar}},
 pages = {332--335},
 publisher = {IEEE},
 isbn = {978-1-4673-6778-3},
 booktitle = {{2015 International Conference on Electrical Engineering and Informatics (ICEEI)}},
 year = {2015}
}

@inproceedings{Leigsnering.2011,
 author = {Leigsnering, Michael and Debes, Christian and Zoubir, Abdelhak M.},
 title = {{Compressive sensing in through-the-wall radar imaging}},
 pages = {4008--4011},
 publisher = {IEEE},
 isbn = {978-1-4577-0538-0},
 booktitle = {{2011 IEEE International Conference on Acoustics, Speech and Signal Processing (ICASSP)}},
 year = {2011}
}

@article{Leigsnering.2014,
 author = {Leigsnering, Michael and Ahmad, Fauzia and Amin, Moeness G. and Zoubir, Abdelhak M.},
 year = {2014},
 title = {{Multipath exploitation in through-the-wall radar imaging using sparse reconstruction}},
 pages = {920--939},
 volume = {50},
 number = {2},
 journal = {{IEEE Transactions on Aerospace and Electronic Systems}}
}

@book{Li.2009,
 year = {2009},
 title = {{MIMO Radar Signal Processing}},
 address = {Hoboken, NJ},
 publisher = {Wiley},
 isbn = {978-0-470-17898-0},
 editor = {Li, Jian and Stoica, Petre}
}

@article{Li.2021b,
 author = {Li, Zhi and Jin, Tian and Dai, Yongpeng and Song, Yongkun},
 year = {2021},
 title = {{Through-Wall Multi-Subject Localization and Vital Signs Monitoring Using UWB MIMO Imaging Radar}},
 pages = {2905},
 volume = {13},
 number = {15},
 journal = {{Remote Sensing}}
}

@article{Lim.2022,
 author = {Lim, Sungmook and Jang, Gwang Soo and Song, Wonyoung and Kim, Baek-Hyun and Kim, Dong Hyun},
 year = {2022},
 title = {{Non-Contact VITAL Signs Monitoring of a Patient Lying on Surgical Bed Using Beamforming FMCW Radar}},
 volume = {22},
 number = {21},
 journal = {{Sensors}}
}

@article{Liu.2014,
 author = {Liu, Lanbo and Liu, Sixin},
 year = {2014},
 title = {{Remote Detection of Human Vital Sign With Stepped-Frequency Continuous Wave Radar}},
 pages = {775--782},
 volume = {7},
 number = {3},
 journal = {{IEEE Journal of Selected Topics in Applied Earth Observations and Remote Sensing}}
}

@article{Liu.2017,
 author = {Liu, Cai and Song, Chao and Lu, Qi},
 year = {2017},
 title = {{Random noise de-noising and direct wave eliminating based on SVD method for ground penetrating radar signals}},
 pages = {125--133},
 volume = {144},
 journal = {{Journal of Applied Geophysics}}
}

@inproceedings{Liu.2020,
 author = {Liu, Jian and Chen, Yingying and Dong, Yudi and Wang, Yan and Zhao, Tiannming and Yao, Yu-Dong},
 title = {{Continuous User Verification via Respiratory Biometrics}},
 pages = {1--10},
 publisher = {IEEE},
 isbn = {978-1-7281-6412-0},
 editor = {Rawat, Danda B. and Wang, Wei and Yang, Dejun},
 booktitle = {{IEEE INFOCOM 2020 - IEEE Conference on Computer Communications}},
 year = {2020}
}

@booklet{Machkour.2021,
 author = {Machkour, Jasin and Muma, Michael and Palomar, Daniel P.},
 year = {2022},
 title = {{The Terminating-Random Experiments Selector: Fast High-Dimensional  Variable Selection with False Discovery Rate Control}}
}

@inproceedings{Meng.2019,
 author = {Meng, Kevin and Meng, Yu},
 title = {{Through-Wall Pose Imaging in Real-Time with a Many-to-Many Encoder/Decoder Paradigm}},
 pages = {14--21},
 publisher = {IEEE},
 isbn = {978-1-7281-4550-1},
 booktitle = {{2019 18th IEEE International Conference On Machine Learning And Applications (ICMLA)}},
 year = {2019}
}

@article{Mercuri.2021,
 author = {Mercuri, Marco and Lu, Yiting and Polito, Salvatore and Wieringa, Fokko and Liu, Yao-Hong and {van der Veen}, Alle-Jan and {van Hoof}, Chris and Torfs, Tom},
 year = {2021},
 title = {{Enabling Robust Radar-Based Localization and Vital Signs Monitoring in Multipath Propagation Environments}},
 pages = {3228--3240},
 volume = {68},
 number = {11},
 journal = {{IEEE Transactions on Biomedical Engineering}}
}

@article{Mercuri.2021b,
 author = {Mercuri, Marco and Sacco, Giulia and Hornung, Rainer and Zhang, Peng and Visser, Hubregt J. and Hijdra, Martijn and Liu, Yao-Hong and Pisa, Stefano and {van Liempd}, Barend and Torfs, Tom},
 year = {2021},
 title = {{2-D Localization, Angular Separation and Vital Signs Monitoring Using a SISO FMCW Radar for Smart Long-Term Health Monitoring Environments}},
 pages = {11065--11077},
 volume = {8},
 number = {14},
 journal = {{IEEE Internet of Things Journal}}
}

@inproceedings{Muma.2012,
 author = {Muma, Michael and Cheng, Yao and Roemer, Florian and Haardt, Martin and Zoubir, Abdelhak M.},
 title = {{Robust source number enumeration for r-dimensional arrays in case of brief sensor failures}},
 pages = {3709--3712},
 publisher = {IEEE},
 isbn = {978-1-4673-0046-9},
 booktitle = {{2012 IEEE International Conference on Acoustics, Speech and Signal Processing (ICASSP)}},
 year = {2012}
}

@article{Nahar.2018b,
 author = {Nahar, Sabikun and Phan, Tuan and Quaiyum, Farhan and Ren, Lingyun and Fathy, Aly E. and Kilic, Ozlem},
 year = {2018},
 title = {{An Electromagnetic Model of Human Vital Signs Detection and Its Experimental Validation}},
 pages = {338--349},
 volume = {8},
 number = {2},
 journal = {{IEEE Journal on Emerging and Selected Topics in Circuits and Systems}}
}

@book{Nguyen.2016,
 author = {Nguyen, Cam and Park, Joongsuk},
 year = {2016},
 title = {{Stepped-Frequency Radar Sensors}},
 address = {Cham},
 publisher = {{Springer International Publishing}},
 isbn = {978-3-319-12270-0}
}

@article{Niu.2020,
 author = {Niu, Haoyu and Wang, Yanan and Zhao, Tiebiao and Chen, YangQuan},
 year = {2020},
 title = {{A Low-cost Soil Moisture Monitoring Method by Using Walabot and Machine Learning Algorithms}},
 pages = {15784--15789},
 volume = {53},
 number = {2},
 journal = {{IFAC-PapersOnLine}}
}

@article{Odendaal.1994,
 author = {Odendaal, J. W. and Barnard, E. and Pistorius, C.W.I.},
 year = {1994},
 title = {{Two-dimensional superresolution radar imaging using the MUSIC algorithm}},
 pages = {1386--1391},
 volume = {42},
 number = {10},
 journal = {{IEEE Transactions on Antennas and Propagation}}
}

@inproceedings{Ogawa.2018,
 author = {Ogawa, K. and Kajiwara, A.},
 title = {{2D high resolution of stepped-FM radar based on MUSIC scheme}},
 pages = {51--54},
 publisher = {IEEE},
 isbn = {978-1-5386-1286-6},
 booktitle = {{2018 IEEE Topical Conference on Wireless Sensors and Sensor Networks (WiSNet)}},
 year = {2018}
}

@article{Paterniani.2023,
 author = {Paterniani, Giacomo and Sgreccia, Daria and Davoli, Alessandro and Guerzoni, Giorgio and {Di Viesti}, Pasquale and Valenti, Anna Chiara and Vitolo, Marco and Vitetta, Giorgio M. and Boriani, Giuseppe},
 year = {2023},
 title = {{Radar-Based Monitoring of Vital Signs: A Tutorial Overview}},
 pages = {277--317},
 volume = {111},
 number = {3},
 journal = {{Proceedings of the IEEE}}
}

@article{Pillai.1989,
 author = {Pillai, S. U. and Kwon, B. H.},
 year = {1989},
 title = {{Forward/backward spatial smoothing techniques for coherent signal identification}},
 pages = {8--15},
 volume = {37},
 number = {1},
 journal = {{IEEE Transactions on Acoustics, Speech, and Signal Processing}}
}

@inproceedings{Quaiyum.2017,
 author = {Quaiyum, Farhan and Ren, Lingyun and Nahar, Sabikun and Foroughian, Farnaz and Fathy, Aly E.},
 title = {{Development of a reconfigurable low cost multi-mode radar system for contactless vital signs detection}},
 pages = {1245--1247},
 publisher = {IEEE},
 isbn = {978-1-5090-6360-4},
 booktitle = {{2017 IEEE MTT-S International Microwave Symposium (IMS)}},
 year = {2017}
}

@booklet{Rao.2018,
 author = {Rao, Sandeep},
 year = {2018},
 title = {{MIMO Radar - Application Report}}
}

@article{Ren.2015,
 author = {Ren, Lingyun and Koo, Yun Seo and Wang, Haofei and Wang, Yazhou and Liu, Quanhua and Fathy, Aly E.},
 year = {2015},
 title = {{Noncontact Multiple Heartbeats Detection and Subject Localization Using UWB Impulse Doppler Radar}},
 pages = {690--692},
 volume = {25},
 number = {10},
 journal = {{IEEE Microwave and Wireless Components Letters}}
}

@inproceedings{Ren.2015b,
 author = {Ren, Lingyun and Wang, Haofei and Naishadham, Krishna and Liu, Quanhua and Fathy, Aly E.},
 title = {{Non-invasive detection of cardiac and respiratory rates from stepped frequency continuous wave radar measurements using the state space method}},
 pages = {1--4},
 publisher = {IEEE},
 isbn = {978-1-4799-8275-2},
 booktitle = {{2015 IEEE MTT-S International Microwave Symposium}},
 year = {2015}
}

@article{Ren.2017,
 author = {Ren, Lingyun and Kong, Lingqin and Foroughian, Farnaz and Wang, Haofei and Theilmann, Paul and Fathy, Aly E.},
 year = {2017},
 title = {{Comparison Study of Noncontact Vital Signs Detection Using a Doppler Stepped-Frequency Continuous-Wave Radar and Camera-Based Imaging Photoplethysmography}},
 pages = {3519--3529},
 volume = {65},
 number = {9},
 journal = {{IEEE Transactions on Microwave Theory and Techniques}}
}

@book{Richards.2005,
 author = {Richards, Mark A.},
 year = {2005},
 title = {{Fundamentals of Radar Signal Processing}},
 address = {New York, NY},
 publisher = {McGraw-Hill},
 isbn = {0071444742},
 series = {{McGraw-Hill electronic engineering}}
}

@article{Schellenberger.2020,
 author = {Schellenberger, Sven and Shi, Kilin and Michler, Fabian and Lurz, Fabian and Weigel, Robert and Koelpin, Alexander},
 year = {2020},
 title = {{Continuous In-Bed Monitoring of Vital Signs Using a Multi Radar Setup for Freely Moving Patients}},
 volume = {20},
 number = {20},
 journal = {{Sensors}}
}

@article{Schroth.2021,
 author = {Schroth, Christian A. and Muma, Michael},
 year = {2021},
 title = {{Robust M-Estimation Based Bayesian Cluster Enumeration for Real Elliptically Symmetric Distributions}},
 pages = {3525--3540},
 volume = {69},
 journal = {{IEEE Transactions on Signal Processing}}
}

@article{Seifert.2019,
 author = {Seifert, Ann-Kathrin and Amin, Moeness G. and Zoubir, Abdelhak M.},
 year = {2019},
 title = {{Toward Unobtrusive In-Home Gait Analysis Based on Radar Micro-Doppler Signatures}},
 pages = {2629--2640},
 volume = {66},
 number = {9},
 journal = {{IEEE Transactions on Biomedical Engineering}}
}

@article{Seo.2021,
 author = {Seo, Jiho and Lee, Jonghyeok and Park, Jaehyun and Kim, Hyungju and You, Sungjin},
 year = {2021},
 title = {{Distributed Two-Dimensional MUSIC for Joint Range and Angle Estimation with Distributed FMCW MIMO Radars}},
 volume = {21},
 number = {22},
 journal = {{Sensors}}
}

@article{Shahimaeen.2019,
 author = {Shahimaeen, Assieh and Dehghani, Mohamad Javad},
 year = {2019},
 title = {{Two--dimensional DOA estimation for coherent signals using a novel covariance--like matrix}},
 volume = {30},
 number = {6},
 journal = {{Transactions on Emerging Telecommunications Technologies}}
}

@article{Shyu.2019,
 author = {Shyu, Kuo-Kai and Chiu, Luan-Jiau and Lee, Po-Lei and Tung, Tzu-Han and Yang, Shun-Han},
 year = {2019},
 title = {{Detection of Breathing and Heart Rates in UWB Radar Sensor Data Using FVPIEF-Based Two-Layer EEMD}},
 pages = {774--784},
 volume = {19},
 number = {2},
 journal = {{IEEE Sensors Journal}}
}

@article{Starr.2014,
 author = {Starr, Joseph W. and Lattimer, Brian Y.},
 year = {2014},
 title = {{Evaluation of Navigation Sensors in Fire Smoke Environments}},
 pages = {1459--1481},
 volume = {50},
 number = {6},
 journal = {{Fire Technology}}
}

@article{Su.2019,
 author = {Su, Wei-Chih and Tang, Mu-Cyun and Arif, Rezki El and Horng, Tzyy-Sheng Jason and Wang, Fu-Kang},
 year = {2019},
 title = {{Stepped-Frequency Continuous-Wave Radar With Self-Injection-Locking Technology for Monitoring Multiple Human Vital Signs}},
 pages = {5396--5405},
 volume = {67},
 number = {12},
 journal = {{IEEE Transactions on Microwave Theory and Techniques}}
}

@article{Takao.1987,
 author = {Takao, K. and Kikuma, N.},
 year = {1987},
 title = {{An adaptive array utilizing an adaptive spatial averaging technique for multipath environments}},
 pages = {1389--1396},
 volume = {35},
 number = {12},
 journal = {{IEEE Transactions on Antennas and Propagation}}
}

@article{Tan.2018,
 author = {Tan, Ming and Wang, Chun-yang and Li, Zhi-hui and Li, Xin and Bao, Lei},
 year = {2018},
 title = {{Stepped Frequency Pulse Frequency Diverse Array Radar for Target Localization in Angle and Range Domains}},
 pages = {1--12},
 volume = {2018},
 journal = {{International Journal of Antennas and Propagation}}
}

@inproceedings{Torchalla.2021,
 author = {Torchalla, Moritz and Schnaubelt, Marius and Daun, Kevin and von Stryk, Oskar},
 title = {{Robust Multisensor Fusion for Reliable Mapping and Navigation in Degraded Visual Conditions}},
 pages = {110--117},
 publisher = {IEEE},
 isbn = {978-1-6654-1764-8},
 booktitle = {{2021 IEEE International Symposium on Safety, Security, and Rescue Robotics (SSRR)}},
 year = {2021}
}

@article{Uthayakumar.2022,
 author = {Uthayakumar, Akileshwaran and Mohan, Manoj Prabhakar and Khoo, Eng Huat and Jimeno, Joe and Siyal, Mohammed Yakoob and Karim, Muhammad Faeyz},
 year = {2022},
 title = {{Machine Learning Models for Enhanced Estimation of Soil Moisture Using Wideband Radar Sensor}},
 pages = {5810},
 volume = {22},
 number = {15},
 journal = {{Sensors}}
}

@article{Wang.2012,
 author = {Wang, Wen-Qin},
 year = {2012},
 title = {{Virtual Antenna Array Analysis for MIMO Synthetic Aperture Radars}},
 pages = {1--10},
 volume = {2012},
 journal = {{International Journal of Antennas and Propagation}}
}

@article{Wang.2014,
 author = {Wang, Jingyu and Wang, Xiang and Chen, Lei and Huangfu, Jiangtao and Li, Changzhi and Ran, Lixin},
 year = {2014},
 title = {{Noncontact Distance and Amplitude-Independent Vibration Measurement Based on an Extended DACM Algorithm}},
 pages = {145--153},
 volume = {63},
 number = {1},
 journal = {{IEEE Transactions on Instrumentation and Measurement}}
}

@article{Wang.2015,
 author = {Wang, Siying and Pohl, Antje and Jaeschke, Timo and Czaplik, Michael and K{\"o}ny, Marcus and Leonhardt, Steffen and Pohl, Nils},
 year = {2015},
 title = {{A novel ultra-wideband 80 GHz FMCW radar system for contactless monitoring of vital signs}},
 pages = {4978--4981},
 volume = {2015},
 journal = {{Annual International Conference of the IEEE Engineering in Medicine and Biology Society}}
}

@article{Wax.1985,
 author = {Wax, M. and Kailath, T.},
 year = {1985},
 title = {{Detection of signals by information theoretic criteria}},
 pages = {387--392},
 volume = {33},
 number = {2},
 journal = {{IEEE Transactions on Acoustics, Speech, and Signal Processing}}
}

@book{Wehner.1995,
 author = {Wehner, Donald R.},
 year = {1995},
 title = {{High-Resolution Radar}},
 address = {Boston},
 publisher = {{Artech House}},
 isbn = {0-89006-727-9}
}

@article{Wu.2023,
 author = {Wu, Qisong and Huang, Xinyue and Chen, Yalong and Li, Jinzhao and Zhu, Wenqiang},
 year = {2023},
 title = {{Multitarget Respiration Monitoring Based on Cumulative Phase Gradient Approach}},
 pages = {1--5},
 volume = {20},
 journal = {{IEEE Geoscience and Remote Sensing Letters}}
}

@inproceedings{Yi.2005,
 author = {Yi, Huiyue and Zhou, Xilang},
 title = {{On 2D forward-backward spatial smoothing for azimuth and elevation estimation of coherent signals}},
 pages = {80--83},
 publisher = {IEEE},
 isbn = {0-7803-8883-6},
 booktitle = {{2005 IEEE Antennas and Propagation Society International Symposium}},
 year = {2005}
}

@article{Zhang.2018,
 author = {Zhang, Rui and Quan, Ying-Hui and Zhu, Sheng-Qi and Yang, Lei and Li, Ya-chao and Xing, Meng-Dao},
 year = {2018},
 title = {{Joint High-Resolution Range and DOA Estimation via MUSIC Method Based on Virtual Two-Dimensional Spatial Smoothing for OFDM Radar}},
 pages = {1--9},
 volume = {2018},
 journal = {{International Journal of Antennas and Propagation}}
}

\onecolumn
\appendices
\section{Table of all Measurements}
\begin{table}[H]
	\begin{center}
		\begin{tabular}{c c c c c c}
            \toprule
            \bfseries ID & \bfseries Obstacle & \bfseries Range in m & \bfseries DoA & \bfseries Posture & \bfseries Person
            \\\hline
            \begin{filecontents*}{figures/resultsTables/data_table_all.csv}
range_1,range_2,range_3,range_4,range_5,doa_1,doa_2,doa_3,doa_4,doa_5,posture_1,posture_2,posture_3,posture_4,posture_5,obstacle,person_ID_1,person_ID_2,person_ID_3,person_ID_4,person_ID_5,ID
,,,,,,,,,,,,,,,F,,,,,,0
,,,,,,,,,,,,,,,W,,,,,,0-1
,,,,,,,,,,,,,,,D,,,,,,0-2
1,,,,,-45,,,,,2,,,,,F,1,,,,,1
2,,,,,-45,,,,,1,,,,,F,1,,,,,2
3,,,,,0,,,,,0,,,,,F,1,,,,,3
3,,,,,0,,,,,0,,,,,D,1,,,,,4
1,,,,,0,,,,,3,,,,,F,2,,,,,5
3.5,,,,,-45,,,,,0,,,,,F,2,,,,,6
2,,,,,0,,,,,4,,,,,F,2,,,,,7
1.5,,,,,-45,,,,,0,,,,,D,2,,,,,8
2.5,,,,,0,,,,,0,,,,,W,1,,,,,9
2.5,,,,,-45,,,,,2,,,,,W,1,,,,,10
1,,,,,0,,,,,0,,,,,W,1,,,,,11
4,2,2,3,3,0,-45,-60,-30,45,0,0,0,0,0,F,1,2,3,4,5,12
2,2,2,2.75,2.5,-60,0,-45,30,45,0,0,0,0,0,F,1,2,3,4,5,13
1.5,2,3,2.75,1,-45,30,-30,45,60,0,0,0,0,1,F,1,2,3,4,5,14
2,1,3.5,2.5,2.5,-45,-30,0,45,60,0,0,1,0,0,F,1,2,3,4,5,15
1.5,2,2,2,3,-60,0,-30,45,30,0,0,0,0,0,F,1,2,3,4,5,16
1.5,1,1.5,2.5,2.5,-60,0,-30,30,45,0,0,0,0,0,F,1,2,3,4,5,17
3,1.5,3,1,2.5,0,45,30,-60,-45,0,0,0,0,0,F,1,2,3,4,5,18
1,3.5,2,2.5,1.5,-60,0,-45,30,45,0,0,0,0,0,F,1,2,3,4,5,19
1,2,2,2.5,3,-60,-30,-45,0,30,0,0,0,1,0,F,1,2,3,4,5,20
1,2,2.5,2.75,2,-60,0,-30,30,45,0,0,0,0,0,F,1,2,3,4,5,21
1,3,3,2,1,-60,30,0,45,60,0,0,0,0,0,F,1,2,3,4,5,22
1,3.5,1.5,2.5,1.5,-60,0,-30,30,60,0,0,0,0,1,F,1,2,3,4,5,23
2.5,1,3,1.5,2,-45,0,-30,30,60,0,0,0,0,0,F,1,2,3,4,5,24
2,2.5,2,1,,-30,30,0,60,,0,0,0,1,,F,1,2,3,4,,25
1,3,2,1.5,,-60,30,0,60,,0,0,1,0,,F,1,2,3,4,,26
3,4,3,,,-30,0,30,,,0,0,0,,,F,1,2,3,,,27
3,4,3,,,-30,0,30,,,0,0,0,,,W,1,2,3,,,27-1
3,4,3,,,-30,0,30,,,0,0,0,,,D,1,2,3,,,27-2
2.5,3,2.5,,,-45,30,-30,,,0,0,0,,,F,1,2,3,,,28
2.5,3,2.5,,,-45,30,-30,,,0,0,0,,,W,1,2,3,,,28-1
2.5,3,2.5,,,-45,30,-30,,,0,0,0,,,D,1,2,3,,,28-2
2,1,2,,,-30,-60,0,,,0,0,0,,,F,1,2,3,,,29
2,1,2,,,-30,-60,0,,,0,0,0,,,W,1,2,3,,,29-1
2,1,2,,,-30,-60,0,,,0,0,0,,,D,1,2,3,,,29-2
2.5,2.5,2.5,,,-45,0,-30,,,0,0,0,,,F,1,2,3,,,30
2.5,2.5,2.5,,,-45,0,-30,,,0,0,0,,,W,1,2,3,,,30-1
2.5,2.5,2.5,,,-45,0,-30,,,0,0,0,,,D,1,2,3,,,30-2
1.25,2.5,2,,,60,45,-30,,,0,0,0,,,F,1,2,3,,,31
1.25,2.5,2,,,60,45,-30,,,0,0,0,,,W,1,2,3,,,31-1
1.25,2.5,2,,,60,45,-30,,,0,0,0,,,D,1,2,3,,,31-2
3,2,2.5,,,0,30,-30,,,0,0,0,,,F,1,2,3,,,32
3,2,2.5,,,0,30,-30,,,0,0,0,,,W,1,2,3,,,32-1
3,2,2.5,,,0,30,-30,,,0,0,0,,,D,1,2,3,,,32-2
1.5,1.5,,,,-30,0,,,,0,0,,,,F,1,3,,,,33
1.5,1.5,,,,-30,0,,,,0,0,,,,W,1,3,,,,33-1
1.5,1.5,,,,-30,0,,,,0,0,,,,D,1,3,,,,33-2
2.5,2.5,,,,-45,-30,,,,0,0,,,,F,2,3,,,,34
2.5,2.5,,,,-45,-30,,,,0,0,,,,W,2,3,,,,34-1
2.5,2.5,,,,-45,-30,,,,0,0,,,,D,2,3,,,,34-2
3,3,,,,-30,0,,,,0,0,,,,F,2,3,,,,35
3,3,,,,-30,0,,,,0,0,,,,W,2,3,,,,35-1
3,3,,,,-30,0,,,,0,0,,,,D,2,3,,,,35-2
3,3,,,,45,-30,,,,0,0,,,,F,1,2,,,,36
3,3,,,,45,-30,,,,0,0,,,,W,1,2,,,,36-1
3,3,,,,30,-30,,,,0,0,,,,D,1,2,,,,36-2
3,1.5,,,,-60,-30,,,,0,0,,,,F,1,2,,,,37
3,1.5,,,,-60,-30,,,,0,0,,,,W,1,2,,,,37-1
3,1.5,,,,-45,-30,,,,0,0,,,,D,1,2,,,,37-2
\end{filecontents*}
\csvreader[range = 1-3,
            ]{figures/resultsTables/data_table_all.csv}{obstacle=\obstacle,
            ID=\ID}{%
            \\ \ID & \obstacle & \{\} & \{\} & \{\} & \{\}
            }%
            \csvreader[range = 4-14
            ]{figures/resultsTables/data_table_all.csv}{range_1=\rangeOne,
            doa_1=\doaOne,
            posture_1=\postureOne,
            obstacle=\obstacle,
            person_ID_1=\personIDOne,
            ID=\ID}{%
            \\ \ID & \obstacle & \{\rangeOne\} & \{\doaOne\} & \{\postureOne\} & \{\personIDOne\}
            }%
            \csvreader[range = 15-27
            ]{figures/resultsTables/data_table_all.csv}{range_1=\rangeOne,
            range_2=\rangeTwo,
            range_3=\rangeThree,
            range_4=\rangeFour,
            range_5=\rangeFive,
            doa_1=\doaOne,
            doa_2=\doaTwo,
            doa_3=\doaThree,
            doa_4=\doaFour,
            doa_5=\doaFive,
            posture_1=\postureOne,
            posture_2=\postureTwo,
            posture_3=\postureThree,
            posture_4=\postureFour,
            posture_5=\postureFive,
            obstacle=\obstacle,
            person_ID_1=\personIDOne,
            person_ID_2=\personIDTwo,
            person_ID_3=\personIDThree,
            person_ID_4=\personIDFour,
            person_ID_5=\personIDFive,
            ID=\ID}{%
            \\ \ID & \obstacle & \{\rangeOne, \rangeTwo, \rangeThree, \rangeFour, \rangeFive\} & \{\doaOne, \doaTwo, \doaThree, \doaFour, \doaFive\} & \{\postureOne, \postureTwo, \postureThree, \postureFour, \postureFive\} & \{\personIDOne, \personIDTwo, \personIDThree, \personIDFour, \personIDFive\}
            }%
            \csvreader[range = 28-29
            ]{figures/resultsTables/data_table_all.csv}{range_1=\rangeOne,
            range_2=\rangeTwo,
            range_3=\rangeThree,
            range_4=\rangeFour,
            doa_1=\doaOne,
            doa_2=\doaTwo,
            doa_3=\doaThree,
            doa_4=\doaFour,
            posture_1=\postureOne,
            posture_2=\postureTwo,
            posture_3=\postureThree,
            posture_4=\postureFour,
            obstacle=\obstacle,
            person_ID_1=\personIDOne,
            person_ID_2=\personIDTwo,
            person_ID_3=\personIDThree,
            person_ID_4=\personIDFour,
            ID=\ID}{%
            \\ \ID & \obstacle & \{\rangeOne, \rangeTwo, \rangeThree, \rangeFour\} & \{\doaOne, \doaTwo, \doaThree, \doaFour\} & \{\postureOne, \postureTwo, \postureThree, \postureFour\} & \{\personIDOne, \personIDTwo, \personIDThree, \personIDFour\}
            }%
            \csvreader[range = 30-47
            ]{figures/resultsTables/data_table_all.csv}{range_1=\rangeOne,
            range_2=\rangeTwo,
            range_3=\rangeThree,
            doa_1=\doaOne,
            doa_2=\doaTwo,
            doa_3=\doaThree,
            posture_1=\postureOne,
            posture_2=\postureTwo,
            posture_3=\postureThree,
            obstacle=\obstacle,
            person_ID_1=\personIDOne,
            person_ID_2=\personIDTwo,
            person_ID_3=\personIDThree,
            ID=\ID}{%
            \\ \ID & \obstacle & \{\rangeOne, \rangeTwo, \rangeThree\} & \{\doaOne, \doaTwo, \doaThree\} & \{\postureOne, \postureTwo, \postureThree\} & \{\personIDOne, \personIDTwo, \personIDThree\}
            }%
            \csvreader[range = 48-62
            ]{figures/resultsTables/data_table_all.csv}{range_1=\rangeOne,
            range_2=\rangeTwo,
            doa_1=\doaOne,
            doa_2=\doaTwo,
            posture_1=\postureOne,
            posture_2=\postureTwo,
            obstacle=\obstacle,
            person_ID_1=\personIDOne,
            person_ID_2=\personIDTwo,
            ID=\ID}{%
            \\ \ID & \obstacle & \{\rangeOne, \rangeTwo\} & \{\doaOne, \doaTwo\} & \{\postureOne, \postureTwo\} & \{\personIDOne, \personIDTwo\}
            }%
            \\\bottomrule
        \end{tabular}
	\end{center}
	\caption{Experiments conducted, Posture: 0: sitting, 1: laying, 2: standing, 3: sitting backwards to radar, 4: sitting sideways to radar, obstacle: F: free space, D: door, W: wall}
	\label{tab:measurements}
\end{table}

\begin{table}
	\centering
	\begin{tabular}{c c c c c} 
		\toprule
		\bfseries ID & \bfseries Obstacle & \bfseries \makecell{Mean Location \\ Error in m} & \bfseries TPP & \bfseries FDP \\ 
            \hline
            \begin{filecontents*}{figures/resultsTables/eval_round.csv}
ID,LastLocError,LastTPP,LastFDP,obstacle
0,  - ,  - ,  - ,F
0-1,  - ,  - ,  - ,W
0-2,  - ,  - ,  - ,D
1,0.16,   1,0.75,F
2,0.24,   1,   0,F
3,0.06,   1, 0.5,F
4,0.11,   1,   0,D
5,0.12,   1, 0.8,F
6, 0.3,   1,   0,F
7,0.15,   1, 0.5,F
8,0.19,   1, 0.5,D
9,  - ,   0,   1,W
10,  - ,   0,  - ,W
11,0.22,   1, 0.8,W
12,0.11, 0.8,0.13,F
13,0.17, 0.8,0.25,F
14,0.15,   1,   0,F
15,0.11, 0.6,0.29,F
16,0.12,   1,   0,F
17,0.11, 0.6,0.29,F
18,0.15, 0.8,0.22,F
19,0.08, 0.8, 0.5,F
20,0.15, 0.6, 0.7,F
21, 0.1,   1,0.25,F
22,0.12, 0.8,0.17,F
23,0.13, 0.8,0.29,F
24,0.13,   1,   0,F
25,0.06,0.75, 0.4,F
26,0.11, 0.5,0.38,F
27,0.08,0.67, 0.5,F
27-1,  - ,   0,  - ,W
27-2,0.11,0.67,0.33,D
28,0.12,   1, 0.4,F
28-1,  - ,   0,   1,W
28-2,0.18,0.67, 0.6,D
29,0.14,   1,0.29,F
29-1,  - ,   0,   1,W
29-2,0.19,0.67, 0.5,D
30, 0.1,0.33,0.75,F
30-1,0.22,0.33, 0.5,W
30-2,0.16,   1, 0.4,D
31, 0.1,   1,0.33,F
31-1,  - ,   0,   1,W
31-2,0.15,0.67, 0.5,D
32,0.09,   1,0.33,F
32-1,  - ,   0,   1,W
32-2,0.16,   1,0.33,D
33,0.08,   1, 0.5,F
33-1,0.13, 0.5,0.67,W
33-2,0.12,   1, 0.5,D
34,0.18, 0.5,0.75,F
34-1,0.27, 0.5,   0,W
34-2,0.06, 0.5,0.75,D
35,0.13,   1, 0.5,F
35-1,  - ,   0,   1,W
35-2,0.23, 0.5, 0.5,D
36,0.17,   1,   0,F
36-1,  - ,   0,  - ,W
36-2,0.25,   1,   0,D
37,0.15, 0.5,   0,F
37-1,  - ,   0,  - ,W
37-2,0.18,   1, 0.5,D
\end{filecontents*}
\csvreader[]{figures/resultsTables/eval_round.csv}{
                LastLocError=\LocError, ID=\ID,LastFDP=\FDP, LastTPP=\TPP, obstacle=\obst}{%
                \\ \ID & \obst &\LocError & \TPP & \FDP}%
		\\\bottomrule
	\end{tabular}
	\caption{Performance metrics for person localization for all experiments.}
	\label{tab:results}
\end{table}

\twocolumn

\end{document}